\documentclass{llncs}
\pdfoutput=1
\usepackage{amsmath}

\usepackage{graphicx}
\usepackage{mathpartir}

\usepackage{cite}

\usepackage{xspace}

\newcommand{\eg}{e.g.,\xspace}
\newcommand{\ie}{i.e.,\xspace}
\newcommand{\Sound}{sound}
\usepackage{thy/document/isabelle,thy/document/isabellesym}

\usepackage{amssymb}

\usepackage{ragged2e}
\usepackage{tabularx} 
\usepackage{booktabs}

\usepackage{longtable}

\usepackage{amsmath}
\usepackage{stmaryrd}
\usepackage{subfig}

\usepackage[defblank]{paralist}
\let\oldinparaenum=\inparaenum
\def\inparaenum{\oldinparaenum[(i)]}

\usepackage{mathpartir}

\usepackage[greek,english]{babel}
\usepackage{type1cm}

\usepackage{tikz}
\usetikzlibrary{calc}

\DeclareRobustCommand{\isascriptstyle}{\def\isamath##1{##1}\def\isatext##1{\mbox{\isastylescript##1}}}

\newcommand{\isabellestylesf}{%
\isabellestyleit%
\renewcommand{\isastyle}{\small\sf}%
\renewcommand{\isastyleminor}{\sf}%
\renewcommand{\isastylescript}{\scriptsize}%
\renewcommand{\isascriptstyle}{\def\isamath####1{####1}\def\isatext####1{\mbox{\isastylescript####1}}\def\isagreek####1{\foreignlanguage{greek}{\mbox{\isastylescript####1}}}}%
\renewcommand{\isacharprime}{\ensuremath{\mskip2mu{'}\mskip-2mu}}%

\DeclareRobustCommand{\isactrlsub}[1]{{\isascriptstyle${}\mathsf{\sb{\vphantom{gb}##1}}$}}%
\DeclareRobustCommand{\isactrlsup}[1]{{\isascriptstyle${}\mathsf{\sp{\vphantom{gb}##1}}$}}%
\DeclareRobustCommand{\isactrlisub}[1]{{\isascriptstyle${}\mathsf{\sb{\vphantom{gb}##1}}$}}%
\DeclareRobustCommand{\isactrlisup}[1]{{\isascriptstyle${}\mathsf{\sp{\vphantom{gb}##1}}$}}%
\DeclareRobustCommand{\isactrlbsub}{\bgroup\isascriptstyle\begin{math}{}\mathsf\bgroup\sb\bgroup}%
\DeclareRobustCommand{\isactrlesub}{\egroup\egroup\end{math}\egroup}%
\DeclareRobustCommand{\isactrlbsup}{\bgroup\isascriptstyle\begin{math}{}\mathsf\bgroup\sp\bgroup}%
\DeclareRobustCommand{\isactrlesup}{\egroup\egroup\end{math}\egroup}%

\renewcommand{\isamarkupheader}[1]{\isastyletext\section{##1}}
\renewcommand{\isamarkupchapter}[1]{\isastyletext\chapter{##1}}
\renewcommand{\isamarkupsection}[1]{\isastyletext\section{##1}}
\renewcommand{\isamarkupsubsection}[1]{\isastyletext\subsection{##1}}
\renewcommand{\isamarkupsubsubsection}[1]{\isastyletext\subsubsection{##1}}
\renewcommand{\isamarkupsect}[1]{\isastyletext\section{##1}}
\renewcommand{\isamarkupsubsect}[1]{\isastyletext\subsection{##1}}
\renewcommand{\isamarkupsubsubsect}[1]{\isastyletext\subsubsection{##1}}

\newcommand{\isagreek}[1]{\foreignlanguage{greek}{\mbox{##1}}}
\renewcommand{\isasymalpha}{\isagreek{a}}
\renewcommand{\isasymbeta}{\isagreek{b}}
\renewcommand{\isasymgamma}{\isagreek{g}}
\renewcommand{\isasymdelta}{\isagreek{d}}
\renewcommand{\isasymepsilon}{\isagreek{e}}
\renewcommand{\isasymzeta}{\isagreek{z}}
\renewcommand{\isasymeta}{\isagreek{h}}
\renewcommand{\isasymtheta}{\isagreek{j}}
\renewcommand{\isasymiota}{\isagreek{i}}
\renewcommand{\isasymkappa}{\isagreek{k}}
\renewcommand{\isasymlambda}{\isamath{\lambda}}
\renewcommand{\isasymmu}{\isagreek{m}}
\renewcommand{\isasymnu}{\isagreek{n}}
\renewcommand{\isasymxi}{\isagreek{x}}
\renewcommand{\isasympi}{\isagreek{p}}
\renewcommand{\isasymrho}{\isagreek{r}}
\renewcommand{\isasymsigma}{\isagreek{sv}}
\renewcommand{\isasymtau}{\isagreek{t}}
\renewcommand{\isasymupsilon}{\isagreek{u}}
\renewcommand{\isasymphi}{\isagreek{f}}
\renewcommand{\isasymchi}{\isagreek{q}}
\renewcommand{\isasympsi}{\isagreek{y}}
\renewcommand{\isasymomega}{\isagreek{w}}
\renewcommand{\isasymGamma}{\isagreek{G}}
\renewcommand{\isasymDelta}{\isagreek{D}}
\renewcommand{\isasymTheta}{\isagreek{J}}
\renewcommand{\isasymLambda}{\isagreek{L}}
\renewcommand{\isasymXi}{\isagreek{X}}
\renewcommand{\isasymPi}{\isagreek{P}}
\renewcommand{\isasymSigma}{\isagreek{Sv}}
\renewcommand{\isasymUpsilon}{\isagreek{U}}
\renewcommand{\isasymPhi}{\isagreek{F}}
\renewcommand{\isasymPsi}{\isagreek{Y}}
\renewcommand{\isasymOmega}{\isagreek{W}}

}
\isabellestyle{sf}

\renewcommand{\isabeginpar}{}
\renewcommand{\isaendpar}{}


\DeclareRobustCommand\ensuretext[1]{\ifmmode\text{#1}\else{#1}\fi}

\newcommand{\freefnt}[1]{\textsl{\rmfamily#1}}
\newcommand{\boundfnt}[1]{{\textsl{\sffamily#1}}}
\newcommand{\constructorfnt}[1]{\textsc{#1}}
\newcommand{\holkeywordfnt}[1]{\texttt{#1}}

\newcommand{\tfreeify}[1]{\ensuretext{\freefnt{#1}}}
\newcommand{\freeify}[1]{\ensuretext{\freefnt{#1}}}
\newcommand{\boundify}[1]{\ensuretext{\boundfnt{#1}}}
\newcommand{\constructor}[1]{\ensuretext{\constructorfnt{#1}}}
\newcommand{\holkeyword}[1]{\ensuretext{\holkeywordfnt{#1}}}

\renewcommand{\isasymturnstile}{\isamath{\,\vdash}}

\renewcommand{\isasymvv}{\mbox{\isastyleminor\isastylescript v+1}}

\newcommand{\Def}[1]{\emph{#1}}

\begin{document}

  \title{A Better Reduction Theorem for Store Buffers}

  \author{Ernie Cohen\inst{1} \and Norbert Schirmer\inst{2}\fnmsep\thanks{Work funded by the German Federal Ministry of Education and Research (BMBF) in the framework of the Verisoft XT project under grant 01 IS 07 008.}}

\institute{Microsoft Corp., Redmond, WA, USA
\and German Research Center for Artificial Intelligence (DFKI) Saarbr\"ucken, Germany\\
\email{ernie.cohen@microsoft.com}, \email{norbert.schirmer@dfki.de}}

\maketitle

\begin{abstract}

When verifying a concurrent program, it is usual to assume that memory
is sequentially consistent.  However, most modern multiprocessors
depend on store buffering for efficiency, and provide native
sequential consistency only at a substantial performance penalty.  To
regain sequential consistency, a programmer has to follow an
appropriate programming discipline. However, na\"ive disciplines,
such as protecting all shared accesses with locks, are not flexible
enough for building high-performance multiprocessor software.

We present a new discipline for concurrent programming under TSO
(total store order, with store buffer forwarding). It does not depend
on concurrency primitives, such as locks. Instead, threads use ghost
operations to acquire and release ownership of memory addresses. A
thread can write to an address only if no other thread owns it, and
can read from an address only if it owns it or it is shared and the thread
has flushed its store buffer since it last wrote to an address it did
not own. This discipline covers both coarse-grained concurrency (where 
data is protected by locks) as well as fine-grained concurrency (where 
atomic operations race to memory).

We formalize this discipline in Isabelle/HOL, and prove that if every
execution of a program in a system without store buffers follows the
discipline, then every execution of the program with store buffers is
sequentially consistent. Thus, we can show sequential consistency
under TSO by ordinary assertional reasoning about the program, without
having to consider store buffers at all.

\end{abstract}

\section{Introduction \label{sec:introduction}}

When verifying a shared-memory concurrent program, it is usual to
assume that each memory operation works directly on a shared memory
state, a model sometimes called \Def{atomic} memory. A memory
implementation that provides this abstraction for programs that
communicate only through shared memory is said to be \Def{sequentially
  consistent}. Concurrent algorithms in the computing literature
tacitly assume sequential consistency, as do most application
programmers.

However, modern computing platforms typically do not guarantee
sequential consistency for arbitrary programs, for two reasons. First,
optimizing compilers are typically incorrect unless the program is
appropriately annotated to indicate which program locations might be
concurrently accessed by other threads; this issue is addressed only
cursorily in this report. Second, modern processors buffer stores of
retired instructions. To make such buffering transparent to
single-processor programs, subsequent reads of the processor read from
these buffers in preference to the cache. (Otherwise, a program could
write a new value to an address but later read an older value.)
However, in a multiprocessor system, processors do not snoop the store
buffers of other processors, so a store is visible to the storing
processor before it is visible to other processors. This can result in
executions that are not sequentially consistent.

The simplest example illustrating such an inconsistency is the
following program, consisting of two threads P0 and P1, where
\texttt{x} and \texttt{y} are shared memory variables (initially 0)
and \texttt{r0} and \texttt{r1} are registers:
\begin{center}
\begin{minipage}{6cm}
\begin{multicols}{3}
P0
\begin{verbatim}
x = 1;
r0 = y;
\end{verbatim}

\columnbreak 

P1
\begin{verbatim}
y = 1;
r1 = x;
\end{verbatim}
\columnbreak 

\end{multicols}
\end{minipage}
\end{center}
In a sequentially consistent execution, it is impossible for both
\texttt{r0} and \texttt{r1} to be assigned $0$. This is because the
assignments to \texttt{x} and \texttt{y} must be executed in some
order; if \texttt{x} (resp. \texttt{y}) is assigned first, then
\texttt{r1} (resp. \texttt{r0}) will be set to $1$. However, in the
presence of store buffers, the assignments to \texttt{r0} and
\texttt{r1} might be performed while the writes to \texttt{x} and
\texttt{y} are still in their respective store buffers, resulting in
both \texttt{r0} and \texttt{r1} being assigned $0$.

One way to cope with store buffers is make them an explicit part of
the programming model. However, this is a substantial programming
concession. First, because store buffers are FIFO, it ratchets up the
complexity of program reasoning considerably; for example, the
reachability problem for a finite set of concurrent finite-state
programs over a finite set of finite-valued locations is in PSPACE
without store buffers, but undecidable (even for two threads) with
store buffers. Second, because writes from function calls might still
be buffered when a function returns, making the store buffers explicit
would break modular program reasoning.

In practice, the usual remedy for store buffering is adherence to a
programming discipline that provides sequential consistency for a
suitable class of architectures. In this report, we describe and prove
the correctness of such a discipline suitable for the memory model
provided by existing x86/x64 machines, where each write emerging from
a store buffer hits a global cache visible to all processors. Because
each processor sees the same global ordering of writes, this model is
sometimes called \Def{total store order}
(TSO)\cite{Adve:Computer-29-12-66}\footnote{Before 2008, Intel
\cite{IntelWhitePaper} and AMD \cite{AMD:AMD64A2006-ALL} both put
forward a weaker memory model in which writes to different memory
addresses may be seen in different orders on different processors, but
respecting causal ordering. However, current implementations satisfy
the stronger conditions described in this report and are also
compliant with the latest revisions of the Intel specifications
\cite{Intel:IIA2006-ALL}. According to Owens et
al. \cite{Owens:TPHOL09-?} AMD is also planning a similar adaptation
of their manuals.}

The concurrency discipline most familiar to concurrent programs is one
where each variable is protected by a lock, and a thread must hold the
corresponding lock to access the variable. (It is possible to
generalize this to allow shared locks, as well as variants such as
split semaphores.) Such lock-based techniques are typically referred
to as \Def{coarse-grained} concurrency control, and suffice for most
concurrent application programming. However, these techniques do not
suffice for low-level system programming (\eg the construction of OS
kernels), for several reasons. First, in kernel programming efficiency
is paramount, and atomic memory operations are more efficient for many
problems. Second, lock-free concurrency control can sometimes
guarantee stronger correctness (\eg wait-free algorithms can provide
bounds on execution time). Third, kernel programming requires taking
into account the implicit concurrency of concurrent hardware
activities (\eg a hardware TLB racing to use page tables while the
kernel is trying to access them), and hardware cannot be forced to
follow a locking discipline.

A more refined concurrency control discipline, one that is much closer
to expert practice, is to classify memory addresses as lock-protected
or shared. Lock-protected addresses are used in the usual way, but
shared addresses can be accessed using atomic operations provided by
hardware (e.g., on x86 class architectures, most reads and writes are
atomic\footnote{This atomicity isn't guaranteed for certain memory types,
  or for operations that cross a cache line.}). The main restriction
on these accesses is that if a processor does a shared write and a subsequent
shared read (possibly from a different address), the processor must
flush the store buffer somewhere in between. For example, in the
example above, both \texttt{x} and \texttt{y} would be shared
addresses, so each processor would have to flush its store buffer
between its first and second operations.

However, even this discipline is not very satisfactory. First, we would
need even more rules to allow locks to be created or destroyed, or to
change memory between shared and protected, and so on.  Second, there
are many interesting concurrency control primitives, and many
algorithms, that allow a thread to obtain exclusive ownership of a
memory address; why should we treat locking as special? 

In this report, we consider a much more general and powerful
discipline that also guarantees sequential consistency. The basic rule
for shared addresses is similar to the discipline above, but there are
no locking primitives. Instead, we treat \Def{ownership} as
fundamental. The difference is that ownership is manipulated by
nonblocking ghost updates, rather than an operation like locking that 
have runtime overhead. Informally the rules of the discipline are as
follows:
\begin{itemize}
\item In any state, each memory address is either \Def{shared} or
  \Def{unshared}. Each memory address is also either \Def{owned} by a
  unique thread or \Def{unowned}. Every unowned address must be
  shared. Each address is also either read-only or read-write. Every
  unshared address must be read-write.
\item A thread can (autonomously) acquire ownership of an unowned
  address, or release ownership of a address that it owns. It can also
  change whether an address it owns is shared or not. Upon release of
  an address it can mark it as read-only.
\item Each memory access is marked as \Def{volatile} or
  \Def{non-volatile}. 
\item A thread can perform a write if it is \Def{\Sound}. It can
  perform a read if it is sound and \Def{clean}.
\item A non-volatile access is \Sound\ if the thread owns the address
  and the address is unshared. A non-volatile read to a read-only
  shared address is also \Sound.
\item A volatile write is \Sound\ if no other thread owns the address
  and the address is not marked as read-only.
\item A volatile read is \Sound\ if the address is shared or the thread owns it.
\item A read is clean if the store buffer has been flushed since the
  last volatile write.  Additionally, a non-volatile read is clean if 
  the store buffer has been flushed since the address was acquired.
\item For interlocked operations (like compare and swap), which have
  the side effect of the store buffer getting flushed, the rules for
  volatile accesses apply.
\end{itemize}

Note first that these conditions are not thread-local, because some actions
are allowed only when an address is unowned, marked read-only, or not
marked read-only. A thread can ascertain such conditions only through
system-wide invariants, respected by all threads, along with data it
reads. By imposing suitable global invariants, various thread-local
disciplines (such as one where addresses are protected by locks,
conditional critical reasons, or monitors) can be derived as lemmas by
ordinary program reasoning, without need for metatheory.

Second, note that these rules can be checked in the context of a
concurrent program without store buffers, by introducing ghost state
to keep track of ownership, whether the thread has performed a write
since the last flush, and which owned addresses were acquired since
the last flush. Our main result is that if a program obeys the rules
above, then the program is sequentially consistent when executed on a
TSO machine.

Consider our first example program. If we choose to leave both
\texttt{x} and \texttt{y} shared, then all accesses must be volatile.
This would force each thread to flush the store buffer between their
first and second operations.  In practice, on an x86/x64 machine, this
would be done by making the writes interlocked, which flushes store
buffers as a side effect.  Whichever thread flushes its store buffer
second is guaranteed to see the write of the other thread, making the
execution violating sequential consistency impossible.

However, couldn't the first thread try to take ownership of \texttt{x}
before writing it, so that its write could be non-volatile? The answer
is that it could, but then the second thread would be unable to read
\texttt{x} volatile (or take ownership of \texttt{x} and read it
non-volatile), because we would be unable to prove that \texttt{x} is
unowned at that point.  In other words, a thread can take ownership of
an address only if it is not racing to do so.

Ultimately, the races allowed by the discipline involve volatile
access to a shared address, which brings us back to locks.  A spinlock
is typically implemented with an interlocked read-modify-write on an
address (the interlocking providing the required flushing of the store
buffer).  If the locking succeeds, we can prove (using for example a
ghost variable giving the ID of the thread taking the lock) that no
other thread holds the lock, and can therefore safely take ownership
of an address ``protected'' by the lock (using the global invariant
that only the lock owner can own the protected address).  Thus, our
discipline subsumes the better-known disciplines governing
coarse-grained concurrency control.

\paragraph{Overview} 
In Section \ref{sec:preliminaries} we introduce preliminaries of
Isabelle/HOL, the theorem prover in which we mechanized our work. In
Section \ref{sec:discipline} we informally describe the programming
discipline and basic ideas of the formalization, which is detailed in
Section \ref{sec:formalization}. Finally we conclude in Section
\ref{sec:conclusion}.

\begin{isabellebody}%
\def\isabellecontext{Preliminaries}%
\isadelimtheory
\endisadelimtheory
\isatagtheory
\endisatagtheory
{\isafoldtheory}%
\isadelimtheory
\endisadelimtheory
\isadelimML
\endisadelimML
\isatagML
\endisatagML
{\isafoldML}%
\isadelimML
\endisadelimML
\isamarkupsection{Preliminaries \label{sec:preliminaries}%
}
\isamarkuptrue%
\begin{isamarkuptext}%
The formalization presented in this papaer is mechanized and checked within the generic interactive theorem prover \emph{Isabelle}\cite{Paulson:IGTP94}. 
Isabelle is called generic as it provides a framework to formalize various \emph{object logics} declared via natural deduction style inference rules.
The object logic that we employ for our formalization is the higher order logic of \emph{Isabelle/HOL}\cite{Nipkow:IHOL02}. 

This article is written using Isabelle's document generation facilities, which guarantees a close correspondence between the presentation and the actual theory files.
We distinguish formal entities typographically from other text. 
We use a sans serif font for types and constants (including functions and predicates), \eg \isa{map}, a slanted serif font for free variables, \eg \isa{\freeify{x}}, and a slanted sans serif font for bound variables, \eg \isa{\boundify{x}}.
Small capitals are used for data type constructors, \eg \isa{\constructor{Foo}}, and type variables have a leading tick, \eg  \isa{\tfreeify{{\isacharprime}a}}. HOL keywords are typeset in type-writer font, \eg \holkeyword{let}. 

To group common premises and to support modular reasoning Isabelle provides \emph{locales}\cite{Ballarin:TYPES03-34,Ballarin:MKM06-31}. 
A locale provides a name for a context of fixed parameters and premises, together with an elaborate infrastructure to define new locales by inheriting and extending other locales, prove theorems within locales and interpret (instantiate) locales. In our formalization we employ this infrastructure to separate the memory system from the programming language semantics. 

The logical and mathematical notions follow the standard notational conventions with a bias towards functional programming. 
We only present the more unconventional parts here. 
We prefer curried function application, \eg \isa{\freeify{f}\ \freeify{a}\ \freeify{b}} instead of \isa{\freeify{f}{\isacharparenleft}\freeify{a}{\isacharcomma}\ \freeify{b}{\isacharparenright}}.
In this setting the latter becomes a function application to \emph{one} argument, which happens to be a pair.

Isabelle/HOL provides a library of standard types like Booleans, natural numbers, integers, total functions, pairs, lists, and sets. Moreover, there are packages to define new data types and records. 
Isabelle allows polymorphic types, \eg \isa{\tfreeify{{\isacharprime}a}\ list} is the list type with type variable \isa{\tfreeify{{\isacharprime}a}}. 
In HOL all functions are total, \eg \isa{nat\ {\isasymRightarrow}\ nat} is a total function on natural numbers. 
A function update is \isa{\freeify{f}{\isacharparenleft}\freeify{y}\ {\isacharcolon}{\isacharequal}\ \freeify{v}{\isacharparenright}\ {\isasymequiv}\ {\isasymlambda}\boundify{x}{\isachardot}\ \holkeyword{if}\ \boundify{x}\ {\isacharequal}\ \freeify{y}\ \holkeyword{then}\ \freeify{v}\ \holkeyword{else}\ \freeify{f}\ \boundify{x}}.
To formalize partial functions the type \isa{\tfreeify{{\isacharprime}a}\ option} is used. 
It is a data type with two constructors, one to inject values of the base type, \eg \isa{{\isasymlfloor}\freeify{x}{\isasymrfloor}}, and the additional element \isa{{\isasymbottom}}. 
A base value can be projected with the function \isa{the}, which is defined by the sole equation \isa{the\ {\isasymlfloor}\freeify{x}{\isasymrfloor}\ {\isacharequal}\ \freeify{x}}. 
Since HOL is a total logic the term \isa{the\ {\isasymbottom}} is still a well-defined yet un(der)specified value. 
Partial functions are usually represented by the type \isa{\tfreeify{{\isacharprime}a}\ {\isasymRightarrow}\ \tfreeify{{\isacharprime}b}\ option}, abbreviated as \isa{\tfreeify{{\isacharprime}a}\ {\isasymrightharpoonup}\ \tfreeify{{\isacharprime}b}}. 
They are commonly used as \emph{maps}. 
We denote the domain of map  \isa{\freeify{m}} by \isa{dom\ \freeify{m}}. 
A map update is written as \isa{\freeify{m}{\isacharparenleft}\freeify{a}\ {\isasymmapsto}\ \freeify{v}{\isacharparenright}}.
We can restrict the domain of a map \isa{\freeify{m}} to a set \isa{\freeify{A}} by \isa{\freeify{m}{\isasymrestriction}\isactrlbsub \freeify{A}\isactrlesub }. 

The syntax and the operations for lists are similar to functional programming languages like ML or Haskell. 
The empty list is \isa{{\isacharbrackleft}{\isacharbrackright}}, with \isa{\freeify{x}\ {\isasymcdot}\ \freeify{xs}} the element \isa{\freeify{x}} is `consed' to the list \isa{\freeify{xs}}.
With \isa{\freeify{xs}\ {\isacharat}\ \freeify{ys}} list \isa{\freeify{ys}} is appended to list \isa{\freeify{xs}}.
With the term \isa{map\ \freeify{f}\ \freeify{xs}} the function \isa{\freeify{f}} is applied to all elements in \isa{\freeify{xs}}. 
The length of a list is \isa{{\isacharbar}\freeify{xs}{\isacharbar}}, the \isa{\freeify{n}}-th element of a list can be selected with \isa{\freeify{xs}\ensuremath{_{[\freeify{n}]}}} and can be updated via \isa{\freeify{xs}{\isacharbrackleft}\freeify{n}\ {\isacharcolon}{\isacharequal}\ \freeify{v}{\isacharbrackright}}. With \isa{dropWhile\ \freeify{P}\ \freeify{xs}} the prefix for which all elements satisfy predicate \isa{\freeify{P}} are dropped from list \isa{\freeify{xs}}.

Sets come along with the standard operations like union, \ie \isa{\freeify{A}\ {\isasymunion}\ \freeify{B}}, membership, \ie \isa{\freeify{x}\ {\isasymin}\ \freeify{A}} and set inversion, \ie \isa{{\isacharminus}\ \freeify{A}}.



Tuples with more than two components are pairs nested to the right.%
\end{isamarkuptext}%
\isamarkuptrue%
\isadelimtheory
\endisadelimtheory
\isatagtheory
\endisatagtheory
{\isafoldtheory}%
\isadelimtheory
\endisadelimtheory
\end{isabellebody}%

%
\begin{isabellebody}%
\def\isabellecontext{Text}%
\isadelimtheory
\endisadelimtheory
\isatagtheory
\endisatagtheory
{\isafoldtheory}%
\isadelimtheory
\endisadelimtheory
\isadelimproof
\endisadelimproof
\isatagproof
\endisatagproof
{\isafoldproof}%
\isadelimproof
\endisadelimproof
\isadelimproof
\endisadelimproof
\isatagproof
\endisatagproof
{\isafoldproof}%
\isadelimproof
\endisadelimproof
\isadelimML
\endisadelimML
\isatagML
\endisatagML
{\isafoldML}%
\isadelimML
\endisadelimML
\isadelimproof
\endisadelimproof
\isatagproof
\endisatagproof
{\isafoldproof}%
\isadelimproof
\endisadelimproof
\isadelimproof
\endisadelimproof
\isatagproof
\endisatagproof
{\isafoldproof}%
\isadelimproof
\endisadelimproof
\isadelimproof
\endisadelimproof
\isatagproof
\endisatagproof
{\isafoldproof}%
\isadelimproof
\endisadelimproof
\isadelimproof
\endisadelimproof
\isatagproof
\endisatagproof
{\isafoldproof}%
\isadelimproof
\endisadelimproof
\isadelimproof
\endisadelimproof
\isatagproof
\endisatagproof
{\isafoldproof}%
\isadelimproof
\endisadelimproof
\isadelimproof
\endisadelimproof
\isatagproof
\endisatagproof
{\isafoldproof}%
\isadelimproof
\endisadelimproof
\isadelimproof
\endisadelimproof
\isatagproof
\endisatagproof
{\isafoldproof}%
\isadelimproof
\endisadelimproof
\isadelimproof
\endisadelimproof
\isatagproof
\endisatagproof
{\isafoldproof}%
\isadelimproof
\endisadelimproof
\isadelimproof
\endisadelimproof
\isatagproof
\endisatagproof
{\isafoldproof}%
\isadelimproof
\endisadelimproof
\isadelimproof
\endisadelimproof
\isatagproof
\endisatagproof
{\isafoldproof}%
\isadelimproof
\endisadelimproof
\isadelimproof
\endisadelimproof
\isatagproof
\endisatagproof
{\isafoldproof}%
\isadelimproof
\endisadelimproof
\isadelimproof
\endisadelimproof
\isatagproof
\endisatagproof
{\isafoldproof}%
\isadelimproof
\endisadelimproof
\isadelimproof
\endisadelimproof
\isatagproof
\endisatagproof
{\isafoldproof}%
\isadelimproof
\endisadelimproof
\isadelimproof
\endisadelimproof
\isatagproof
\endisatagproof
{\isafoldproof}%
\isadelimproof
\endisadelimproof
\isadelimproof
\endisadelimproof
\isatagproof
\endisatagproof
{\isafoldproof}%
\isadelimproof
\endisadelimproof
\isadelimproof
\endisadelimproof
\isatagproof
\endisatagproof
{\isafoldproof}%
\isadelimproof
\endisadelimproof
\isadelimproof
\endisadelimproof
\isatagproof
\endisatagproof
{\isafoldproof}%
\isadelimproof
\endisadelimproof
\isadelimproof
\endisadelimproof
\isatagproof
\endisatagproof
{\isafoldproof}%
\isadelimproof
\endisadelimproof
\isadelimproof
\endisadelimproof
\isatagproof
\endisatagproof
{\isafoldproof}%
\isadelimproof
\endisadelimproof
\isadelimproof
\endisadelimproof
\isatagproof
\endisatagproof
{\isafoldproof}%
\isadelimproof
\endisadelimproof
\isadelimproof
\endisadelimproof
\isatagproof
\endisatagproof
{\isafoldproof}%
\isadelimproof
\endisadelimproof
\isadelimproof
\endisadelimproof
\isatagproof
\endisatagproof
{\isafoldproof}%
\isadelimproof
\endisadelimproof
\isadelimproof
\endisadelimproof
\isatagproof
\endisatagproof
{\isafoldproof}%
\isadelimproof
\endisadelimproof
\isadelimproof
\endisadelimproof
\isatagproof
\endisatagproof
{\isafoldproof}%
\isadelimproof
\endisadelimproof
\isadelimproof
\endisadelimproof
\isatagproof
\endisatagproof
{\isafoldproof}%
\isadelimproof
\endisadelimproof
\isadelimproof
\endisadelimproof
\isatagproof
\endisatagproof
{\isafoldproof}%
\isadelimproof
\endisadelimproof
\isamarkupsection{Programming discipline \label{sec:discipline}%
}
\isamarkuptrue%
\begin{isamarkuptext}%
For sequential code on a single processor the store buffer is invisible, since reads respect outstanding writes in the buffer. 
This argument can be extended to thread local memory in the context of a multiprocessor architecture. 
Memory typically becomes temporarily thread local by means of locking. 
The C-idiom to identify shared portions of the memory is the \texttt{volatile} 
tag on variables and type declarations. 
Thread local memory can be accessed non-volatilely, whereas accesses to shared memory are tagged as volatile. 
This prevents the compiler from applying certain optimizations to those accesses which could cause undesired behavior, \eg to store intermediate values in registers instead of writing them to the memory.

The basic idea behind the programming discipline is, that before gathering new information about the shared state (via reading) the thread has to make the outstanding changes to the shared state visible to others (by flushing the store buffer). 
This allows to sequentialize the reads and writes to obtain a sequentially consistent execution of the global system. 
In this sequentialization a write to shared memory happens when the write instruction exits the store buffer, and a read from the shared memory happens when all preceding writes have exited.

We distinguish thread local and shared memory by an ownership model. 
Ownership is maintained in ghost state and can be transferred as side effect of write operations and by special ghost operations.
Every thread has a set of owned addresses. Owned addresses of different threads are disjoint. 
Moreover, there is a global set of shared addresses which can additionally be marked as read-only. 
Unowned addresses --- addresses owned by no thread --- can be accessed concurrently by all threads. They are a subset of the shared addresses. The read-only addresses are a subset of the unowned addresses.
We only allow a thread to write to owned addresses and unowned, read-write addresses.
We only allow a thread to read from owned addresses and from shared addresses (even if they are owned by another thread).

All writes to shared memory have to be volatile. Reads from shared addresses also have to be volatile, except if the address is owned (\ie single writer, multiple readers) or if the address is read-only. Moreover, non-volatile writes are restricted to owned, unshared memory.
As long as a thread owns an address it is guaranteed that it is the only one writing to that address. Hence this thread can safely perform non-volatile reads to that address without missing any write. Similar it is safe for any thread to access read-only memory via non-volatile reads since there are no outstanding writes at all.

Recall that a read is \Def{clean} if it is guaranteed that there is no outstanding volatile write (to any address) in the store buffer. Additionally non-volatile reads which where not freshly acquired since the last flush are considered clean.
To regain sequential consistency under the presence of store buffers every thread has to make sure that every read is clean, by flushing the store buffer when necessary. To check the flushing policy of a thread, we keep track of clean reads by means of ghost state. For every thread we maintain a dirty flag and a set of acquired addresses. Both are reset as the store buffer gets flushed. Upon a volatile write the dirty flag is set and as an address is acquired (by ghost operations) this is recorded. The dirty flag and the set of acquired addresses is considered to guarantee that a read is clean.

Table \ref{tab:access-grid} summarizes the access policy and Table \ref{tab:flushing} the associated flushing policy of the programming discipline.
The key motivation is to improve performance by minimizing the number of store buffer flushes, 
while staying sequentially consistent.
The need for flushing the store buffer decreases from interlocked accesses (where flushing is a side-effect) over volatile accesses to non-volatile accesses. From the viewpoint of access rights there is no difference between interlocked and volatile accesses. However, keep in mind that some interlocked operations can read from, modify and write to an address in a single atomic step of the underlying hardware and are typically used in lock-free algorithms or for the implementation of locks.

\begin{table}
\centering
\caption{Programming discipline.}
\captionsetup[table]{position=top}
\captionsetup[subtable]{position=top}
\newcommand{\mycomment}[1]{}
\subfloat[Access policy\label{tab:access-grid}]{
\begin{tabular}{m{1.2cm}@ {\hspace{2mm}}m{1.7cm}@ {\hspace{3mm}}m{1.8cm}m{2.2 cm}}
\toprule
               &     shared        &   shared   &  unshared            \\
               &     (read-write)  &   (read-only)   &                 \\
\midrule
owned          & \mycomment{iRW, iR, iW,} vR, vW, R      & unreachable            &  \mycomment{iRW, iR, iW,} vR, vW, R, W \\
owned \mbox{by other} & \mycomment{iR,}       vR                & unreachable            &                           \\
unowned        & \mycomment{iRW, iR, iW,} vR, vW         & \mycomment{iR,} vR, R              &  unreachable\\
\bottomrule
\multicolumn{4}{l}{(v)olatile, (R)ead, (W)rite}\\
\multicolumn{4}{l}{all reads have to be clean }
\end{tabular}
}\hspace{0.3cm}
%
%
%
\subfloat[Flushing policy\label{tab:flushing}]{
\begin{tabular}{lc}
\toprule
                & flush (before)          \\     
\midrule
interlocked     & as side effect                 \\
vR, R            & if not clean               \\
vW, W           & never                  \\
\bottomrule
\end{tabular}
}

\end{table}%
\end{isamarkuptext}%
\isamarkuptrue%
\isamarkupsection{Formalization \label{sec:formalization}%
}
\isamarkuptrue%
\begin{isamarkuptext}%
In this section we go into the details of our formalization. In our model, we distinguish the plain `memory system' from the 
`programming language semantics' which we both describe as a small-step transition relation. 
During the computation of the programming language memory instructions (read / write) are issued to the memory system, 
which itself returns the results in temporary registers. 
This clean interface allows us to parameterize the program semantics over the 
memory system. Our main theorem allows us to simulate a computation step in the semantics based on a 
memory system with store buffers by \isa{\freeify{n}} steps in the semantics based on a   
sequentially consistent memory system. 
We refer to the former one as \Def{store buffer machine}  and to the latter one as \Def{virtual machine}. The simulation theorem is independent of the programming language.

We continue with introducing the common parts of both machines. In Section \ref{sec:virtualmachine} we then describe the virtual machine and in Section \ref{sec:storebuffermachine} the store buffer machine. Section \ref{sec:couplingrelation} gives some details of our coupling relation which is used for the simulation proof presented in Section \ref{sec:simulation}. Finally, in Section \ref{sec:pimp} we illustrate the integration of a programming language on top of the memory system, by presenting PIMP, a concurrent variant of a WHILE language. 

\medskip
Addresses \isa{\freeify{a}}, values \isa{\freeify{v}} and temporaries \isa{\freeify{t}} are natural numbers. 
Ghost annotations for manipulating the ownership information are the following sets of addresses: the acquired addresses \isa{\freeify{A}}, the unshared (local) fraction \isa{\freeify{L}} of the acquired addresses, the released addresses \isa{\freeify{R}} and the writable fraction \isa{\freeify{W}} of the released addresses (the remaining addresses are considered read-only). 
These ownership annotations are considered as side-effects on volatile writes and interlocked operations (in case a write is performed). 
Moreover, a special ghost instruction allows to acquire addresses.
The possible status changes of an address due to these ownership transfer operations are depicted in Figure \ref{fig:ownership-transfer}.
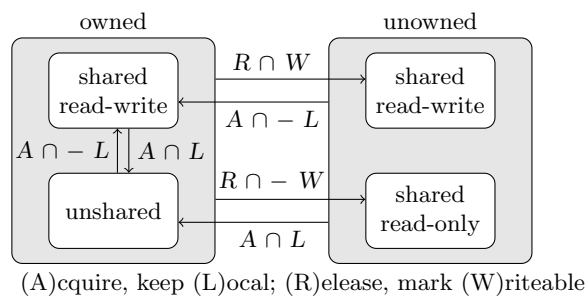
\begin{figure}
\begin{center}
\begin{tikzpicture}
[auto,
 outernode/.style    = {rectangle, rounded corners, draw, text centered, minimum height=3cm, minimum width=2.7cm, fill=gray!20},
 innernode/.style  = {rectangle, rounded corners, draw, text centered, minimum height=1cm, minimum width=1cm, text width=1.5cm, fill=white}
]
\node[outernode] (owned) {};
\node[innernode] (oshared)  [below] at ($ (owned.north) -(0,0.2cm) $) {shared read-write};
\node[innernode] (onshared) [above] at ($ (owned.south) +(0,0.2cm) $) {unshared};
\node[above] at (owned.north) {owned};

\node[outernode] (unowned) [right] at ($ (owned.east) +(1.5cm,0cm) $) {};
\node[innernode] (rwshared)  [below] at ($ (unowned.north) -(0,0.2cm) $) {shared read-write};
\node[innernode] (roshared) [above] at ($ (unowned.south) +(0,0.2cm) $) {shared read-only};
\node[above] at (unowned.north) {unowned};

\path (rwshared.east) -- coordinate (middlex) (oshared.west);

\draw[->] (owned.east |- rwshared.170) -- (rwshared.170); 
  \node [above] at (rwshared.170 -| middlex) {\isa{\freeify{R}\ {\isasyminter}\ \freeify{W}}};

\draw[->] (unowned.west |- oshared.350) -- (oshared.350);
  \node [below] at (oshared.350 -| middlex) {\isa{\freeify{A}\ {\isasyminter}\ {\isacharminus}\ \freeify{L}}};

\draw[->] (unowned.west |- onshared.350) -- (onshared.350);
  \node [below] at (onshared.350 -| middlex) {\isa{\freeify{A}\ {\isasyminter}\ \freeify{L}}};

\draw[->] (owned.east |- roshared.170) -- (roshared.170);
  \node [above] at (roshared.170 -| middlex) {\isa{\freeify{R}\ {\isasyminter}\ {\isacharminus}\ \freeify{W}}};

\draw[->] (oshared.292) -- node {\isa{\freeify{A}\ {\isasyminter}\ \freeify{L}}} (onshared.68);
\draw[->] (onshared.84) -- node {\isa{\freeify{A}\ {\isasyminter}\ {\isacharminus}\ \freeify{L}}} (oshared.276);

\node (legende) [below right] at (owned.south west) {(A)cquire, keep (L)ocal; (R)elease, mark (W)riteable };
\end{tikzpicture}
\end{center}

\caption{Ownership transfer \label{fig:ownership-transfer}}
\end{figure}
A memory instruction is a datatype with the following constructors:
\begin{itemize}
\item \isa{\constructor{Read}\ \freeify{volatile}\ \freeify{a}\ \freeify{t}} for reading from address \isa{\freeify{a}} to temporary \isa{\freeify{t}}, where the Boolean \isa{\freeify{volatile}} determines whether the access is volatile or not.
\item \isa{\constructor{Write}\ \freeify{volatile}\ \freeify{a}\ \freeify{sop}\ \freeify{A}\ \freeify{L}\ \freeify{R}\ \freeify{W}} to write the result of evaluating the store operation \isa{\freeify{sop}} at address \isa{\freeify{a}}. A store operation is a pair \isa{{\isacharparenleft}\freeify{D}{\isacharcomma}\ \freeify{f}{\isacharparenright}}, with the domain \isa{\freeify{D}} and the function \isa{\freeify{f}}.
The function \isa{\freeify{f}} takes temporaries \isa{\freeify{{\isasymtheta}}} as a parameter, which maps a temporary to a value. 
The subset of temporaries that is considered by function \isa{\freeify{f}} is specified by the domain \isa{\freeify{D}}.
We consider store operations as valid when they only depend on their domain: 
\begin{isabelle}%
valid{\isacharunderscore}sop\ \freeify{sop}\ {\isasymequiv}\ {\isasymforall}\boundify{D}\ \boundify{f}\ \boundify{{\isasymtheta}}{\isachardot}\ \freeify{sop}\ {\isacharequal}\ {\isacharparenleft}\boundify{D}{\isacharcomma}\ \boundify{f}{\isacharparenright}\ {\isasymand}\ \boundify{D}\ {\isasymsubseteq}\ dom\ \boundify{{\isasymtheta}}\ {\isasymlongrightarrow}\ \boundify{f}\ \boundify{{\isasymtheta}}\ {\isacharequal}\ \boundify{f}\ {\isacharparenleft}\boundify{{\isasymtheta}}{\isasymrestriction}\isactrlbsub \boundify{D}\isactrlesub {\isacharparenright}%
\end{isabelle}
Again the Boolean \isa{\freeify{volatile}} specifies the kind of memory access.
\item \isa{\constructor{RMW}\ \freeify{a}\ \freeify{t}\ \freeify{sop}\ \freeify{cond}\ \freeify{ret}\ \freeify{A}\ \freeify{L}\ \freeify{R}\ \freeify{W}}, for atomic interlocked `read-modify-write' instructions (flushing the store buffer). First the value at address \isa{\freeify{a}} is loaded to temporary \isa{\freeify{t}}, and then the condition \isa{\freeify{cond}} on the temporaries is considered to decide whether a store operation is also executed. In case of a store the function \isa{\freeify{ret}}, depending on both the old value at address \isa{\freeify{a}} and the new value (according to store operation \isa{\freeify{sop}}), specifies the final result stored in temporary \isa{\freeify{t}}. With a trivial condition \isa{\freeify{cond}} this instruction also covers interlocked reads and writes.
\item \isa{\constructor{Fence}}, a memory fence that flushes the store buffer. 
\item \isa{\constructor{Ghost}\ \freeify{A}\ \freeify{L}} to acquire ownership on addresses.
\end{itemize}

The configuration of a single thread is a tuple \isa{{\isacharparenleft}\freeify{p}{\isacharcomma}\ \freeify{is}{\isacharcomma}\ \freeify{{\isasymtheta}}{\isacharcomma}\ \freeify{sb}{\isacharcomma}\ \freeify{{\isasymD}}{\isacharcomma}\ \freeify{{\isasymO}}{\isacharcomma}\ \freeify{{\isasymA}}{\isacharparenright}} consisting of the program state \isa{\freeify{p}}, a memory instruction list \isa{\freeify{is}}, the map of temporaries \isa{\freeify{{\isasymtheta}}}, the store buffer \isa{\freeify{sb}}, a dirty flag \isa{\freeify{{\isasymD}}} indicating whether there may be an outstanding volatile write in the store buffer, the set of owned addresses \isa{\freeify{{\isasymO}}} and finally the set of addresses \isa{\freeify{{\isasymA}}} acquired since the last store buffer flush. 
The dirty flag \isa{\freeify{{\isasymD}}} and the set \isa{\freeify{{\isasymA}}} are considered to specify if a read is clean: for \emph{all} volatile reads and the non-volatile reads to addresses in \isa{\freeify{{\isasymA}}} the dirty flag must not be set.

The type of the program state \isa{\freeify{p}} and the store buffer \isa{\freeify{sb}} is free. 
For example we later instantiate the store buffer with the union type in case of the virtual machine or with a list of store buffer instructions in case of the machine with store buffer. 

A global configuration \isa{{\isacharparenleft}\freeify{ts}{\isacharcomma}\ \freeify{{\isasymS}}{\isacharcomma}\ \freeify{m}{\isacharparenright}} consists of a list of thread configurations \isa{\freeify{ts}}, a Boolean map  of shared addresses \isa{\freeify{{\isasymS}}} (indicating write permission) and the memory \isa{\freeify{m}}, which is a function from addresses to values. Addresses in the domain of mapping \isa{\freeify{{\isasymS}}} are considered shared and the set of read-only addresses is obtained  from \isa{\freeify{S}} by: \isa{read{\isacharunderscore}only\ \freeify{{\isasymS}}\ {\isasymequiv}\ {\isacharbraceleft}\boundify{a}{\isachardot}\ \freeify{{\isasymS}}\ \boundify{a}\ {\isacharequal}\ {\isasymlfloor}False{\isasymrfloor}{\isacharbraceright}}%
\end{isamarkuptext}%
\isamarkuptrue%
\begin{isamarkuptext}%
We describe the computation of the global system by the non-deterministic transition relation \isa{{\isacharparenleft}\freeify{ts}{\isacharcomma}\ \freeify{{\isasymS}}{\isacharcomma}\ \freeify{m}{\isacharparenright}\ {\isasymRightarrow}\ {\isacharparenleft}\freeify{ts{\isacharprime}}{\isacharcomma}\ \freeify{{\isasymS}{\isacharprime}}{\isacharcomma}\ \freeify{m{\isacharprime}}{\isacharparenright}} defined in Figure~\ref{fig:global-transitions}. 
\begin{figure}
\begin{center}
\isa{\mbox{}\inferrule{\mbox{\freeify{i}\ {\isacharless}\ {\isacharbar}\freeify{ts}{\isacharbar}}\\\ \mbox{\freeify{ts}\ensuremath{_{[\freeify{i}]}}\ {\isacharequal}\ {\isacharparenleft}\freeify{p}{\isacharcomma}\ \freeify{is}{\isacharcomma}\ \freeify{{\isasymtheta}}{\isacharcomma}\ \freeify{sb}{\isacharcomma}\ \freeify{{\isasymD}}{\isacharcomma}\ \freeify{{\isasymO}}{\isacharcomma}\ \freeify{{\isasymA}}{\isacharparenright}}\\\ \mbox{\freeify{{\isasymtheta}}{\isasymturnstile}\ \freeify{p}\ {\isasymrightarrow}\isactrlsub p\ {\isacharparenleft}\freeify{p{\isacharprime}}{\isacharcomma}\ \freeify{is{\isacharprime}}{\isacharparenright}}}{\mbox{{\isacharparenleft}\freeify{ts}{\isacharcomma}\ \freeify{{\isasymS}}{\isacharcomma}\ \freeify{m}{\isacharparenright}\ {\isasymRightarrow}\ {\isacharparenleft}\freeify{ts}{\isacharbrackleft}\freeify{i}\ {\isacharcolon}{\isacharequal}\ {\isacharparenleft}\freeify{p{\isacharprime}}{\isacharcomma}\ \freeify{is}\ {\isacharat}\ \freeify{is{\isacharprime}}{\isacharcomma}\ \freeify{{\isasymtheta}}{\isacharcomma}\ \freeify{record}\ \freeify{p}\ \freeify{p{\isacharprime}}\ \freeify{is{\isacharprime}}\ \freeify{sb}{\isacharcomma}\ \freeify{{\isasymD}}{\isacharcomma}\ \freeify{{\isasymO}}{\isacharcomma}\ \freeify{{\isasymA}}{\isacharparenright}{\isacharbrackright}{\isacharcomma}\ \freeify{{\isasymS}}{\isacharcomma}\ \freeify{m}{\isacharparenright}}}}\\[0.5\baselineskip]
\isa{\mbox{}\inferrule{\mbox{\freeify{i}\ {\isacharless}\ {\isacharbar}\freeify{ts}{\isacharbar}}\\\ \mbox{\freeify{ts}\ensuremath{_{[\freeify{i}]}}\ {\isacharequal}\ {\isacharparenleft}\freeify{p}{\isacharcomma}\ \freeify{is}{\isacharcomma}\ \freeify{{\isasymtheta}}{\isacharcomma}\ \freeify{sb}{\isacharcomma}\ \freeify{{\isasymD}}{\isacharcomma}\ \freeify{{\isasymO}}{\isacharcomma}\ \freeify{{\isasymA}}{\isacharparenright}}\\\ \mbox{{\isacharparenleft}\freeify{is}{\isacharcomma}\ \freeify{{\isasymtheta}}{\isacharcomma}\ \freeify{sb}{\isacharcomma}\ \freeify{m}{\isacharcomma}\ \freeify{{\isasymD}}{\isacharcomma}\ \freeify{{\isasymO}}{\isacharcomma}\ \freeify{{\isasymA}}{\isacharcomma}\ \freeify{{\isasymS}}{\isacharparenright}\ {\isasymrightarrow}\isactrlsub m\ {\isacharparenleft}\freeify{is{\isacharprime}}{\isacharcomma}\ \freeify{{\isasymtheta}{\isacharprime}}{\isacharcomma}\ \freeify{sb{\isacharprime}}{\isacharcomma}\ \freeify{m{\isacharprime}}{\isacharcomma}\ \freeify{{\isasymD}{\isacharprime}}{\isacharcomma}\ \freeify{{\isasymO}{\isacharprime}}{\isacharcomma}\ \freeify{{\isasymA}{\isacharprime}}{\isacharcomma}\ \freeify{{\isasymS}{\isacharprime}}{\isacharparenright}}}{\mbox{{\isacharparenleft}\freeify{ts}{\isacharcomma}\ \freeify{{\isasymS}}{\isacharcomma}\ \freeify{m}{\isacharparenright}\ {\isasymRightarrow}\ {\isacharparenleft}\freeify{ts}{\isacharbrackleft}\freeify{i}\ {\isacharcolon}{\isacharequal}\ {\isacharparenleft}\freeify{p}{\isacharcomma}\ \freeify{is{\isacharprime}}{\isacharcomma}\ \freeify{{\isasymtheta}{\isacharprime}}{\isacharcomma}\ \freeify{sb{\isacharprime}}{\isacharcomma}\ \freeify{{\isasymD}{\isacharprime}}{\isacharcomma}\ \freeify{{\isasymO}{\isacharprime}}{\isacharcomma}\ \freeify{{\isasymA}{\isacharprime}}{\isacharparenright}{\isacharbrackright}{\isacharcomma}\ \freeify{{\isasymS}{\isacharprime}}{\isacharcomma}\ \freeify{m{\isacharprime}}{\isacharparenright}}}}\\[0.5\baselineskip]
\isa{\mbox{}\inferrule{\mbox{\freeify{i}\ {\isacharless}\ {\isacharbar}\freeify{ts}{\isacharbar}}\\\ \mbox{\freeify{ts}\ensuremath{_{[\freeify{i}]}}\ {\isacharequal}\ {\isacharparenleft}\freeify{p}{\isacharcomma}\ \freeify{is}{\isacharcomma}\ \freeify{{\isasymtheta}}{\isacharcomma}\ \freeify{sb}{\isacharcomma}\ \freeify{{\isasymD}}{\isacharcomma}\ \freeify{{\isasymO}}{\isacharcomma}\ \freeify{{\isasymA}}{\isacharparenright}}\\\ \mbox{{\isacharparenleft}\freeify{m}{\isacharcomma}\ \freeify{sb}{\isacharcomma}\ \freeify{{\isasymO}}{\isacharcomma}\ \freeify{{\isasymA}}{\isacharcomma}\ \freeify{{\isasymS}}{\isacharparenright}\ {\isasymrightarrow}\isactrlsub s\isactrlsub b\ {\isacharparenleft}\freeify{m{\isacharprime}}{\isacharcomma}\ \freeify{sb{\isacharprime}}{\isacharcomma}\ \freeify{{\isasymO}{\isacharprime}}{\isacharcomma}\ \freeify{{\isasymA}{\isacharprime}}{\isacharcomma}\ \freeify{{\isasymS}{\isacharprime}}{\isacharparenright}}}{\mbox{{\isacharparenleft}\freeify{ts}{\isacharcomma}\ \freeify{{\isasymS}}{\isacharcomma}\ \freeify{m}{\isacharparenright}\ {\isasymRightarrow}\ {\isacharparenleft}\freeify{ts}{\isacharbrackleft}\freeify{i}\ {\isacharcolon}{\isacharequal}\ {\isacharparenleft}\freeify{p}{\isacharcomma}\ \freeify{is}{\isacharcomma}\ \freeify{{\isasymtheta}}{\isacharcomma}\ \freeify{sb{\isacharprime}}{\isacharcomma}\ \freeify{{\isasymD}}{\isacharcomma}\ \freeify{{\isasymO}{\isacharprime}}{\isacharcomma}\ \freeify{{\isasymA}{\isacharprime}}{\isacharparenright}{\isacharbrackright}{\isacharcomma}\ \freeify{{\isasymS}{\isacharprime}}{\isacharcomma}\ \freeify{m{\isacharprime}}{\isacharparenright}}}}
\end{center}
\caption{Global transitions \label{fig:global-transitions}}
\end{figure}
A transition selects a thread \isa{\freeify{ts}\ensuremath{_{[\freeify{i}]}}\ {\isacharequal}\ {\isacharparenleft}\freeify{p}{\isacharcomma}\ \freeify{is}{\isacharcomma}\ \freeify{{\isasymtheta}}{\isacharcomma}\ \freeify{sb}{\isacharcomma}\ \freeify{{\isasymD}}{\isacharcomma}\ \freeify{{\isasymO}}{\isacharcomma}\ \freeify{{\isasymA}}{\isacharparenright}} and either the `program' the `memory'  or the `store buffer' makes a step. 
These three sub-relations are parameters to the global transition relation. 
The ownership information stored in the ghost components \isa{\freeify{{\isasymD}}}, \isa{\freeify{{\isasymO}}}, \isa{\freeify{{\isasymA}}}, and \isa{\freeify{{\isasymS}}} is sometimes grouped as a single component \isa{\freeify{{\isasymG}}} in the transition rules for succinct presentation.

A program step \isa{\freeify{{\isasymtheta}}{\isasymturnstile}\ \freeify{p}\ {\isasymrightarrow}\isactrlsub p\ {\isacharparenleft}\freeify{p{\isacharprime}}{\isacharcomma}\ \freeify{is{\isacharprime}}{\isacharparenright}} takes the temporaries \isa{\freeify{{\isasymtheta}}} and the current program state \isa{\freeify{p}} and makes a step by returning a new program state \isa{\freeify{p{\isacharprime}}} and an instruction list \isa{\freeify{is{\isacharprime}}} which is appended to the remaining instructions. 
With the functional parameter \isa{\freeify{record}} we are able to maintain bookkeeping information about the program step within the store buffer. It takes the program states \isa{\freeify{p}} and \isa{\freeify{p{\isacharprime}}}, the issued instructions \isa{\freeify{is{\isacharprime}}} and the store buffer \isa{\freeify{sb}} as a parameter. This is a technical device in our proof which allows us to remember program steps of the store buffer machine that are still pending in the virtual machine. 

A memory step \isa{{\isacharparenleft}\freeify{is}{\isacharcomma}\ \freeify{{\isasymtheta}}{\isacharcomma}\ \freeify{sb}{\isacharcomma}\ \freeify{m}{\isacharcomma}\ \freeify{{\isasymD}}{\isacharcomma}\ \freeify{{\isasymO}}{\isacharcomma}\ \freeify{{\isasymA}}{\isacharcomma}\ \freeify{{\isasymS}}{\isacharparenright}\ {\isasymrightarrow}\isactrlsub m\ {\isacharparenleft}\freeify{is{\isacharprime}}{\isacharcomma}\ \freeify{{\isasymtheta}{\isacharprime}}{\isacharcomma}\ \freeify{sb{\isacharprime}}{\isacharcomma}\ \freeify{m{\isacharprime}}{\isacharcomma}\ \freeify{{\isasymD}{\isacharprime}}{\isacharcomma}\ \freeify{{\isasymO}{\isacharprime}}{\isacharcomma}\ \freeify{{\isasymA}{\isacharprime}}{\isacharcomma}\ \freeify{{\isasymS}{\isacharprime}}{\isacharparenright}} of a machine with store buffer may only fill its store buffer.

In a store buffer step \isa{{\isacharparenleft}\freeify{m}{\isacharcomma}\ \freeify{sb}{\isacharcomma}\ \freeify{{\isasymO}}{\isacharcomma}\ \freeify{{\isasymA}}{\isacharcomma}\ \freeify{{\isasymS}}{\isacharparenright}\ {\isasymrightarrow}\isactrlsub s\isactrlsub b\ {\isacharparenleft}\freeify{m{\isacharprime}}{\isacharcomma}\ \freeify{sb{\isacharprime}}{\isacharcomma}\ \freeify{{\isasymO}{\isacharprime}}{\isacharcomma}\ \freeify{{\isasymA}{\isacharprime}}{\isacharcomma}\ \freeify{{\isasymS}{\isacharprime}}{\isacharparenright}} the store buffer may release outstanding instructions to the memory.%
\end{isamarkuptext}%
\isamarkuptrue%
\isamarkupsubsection{Virtual machine \label{sec:virtualmachine}%
}
\isamarkuptrue%
\begin{isamarkuptext}%
The virtual machine is a sequentially consistent machine without store buffers.
The transition rules for its memory system are defined in Figure~\ref{fig:direct-memory}. The store buffer, which is irrelevant in this transition system is referenced by \isa{\freeify{x}}. We instantiate the global transition system with these rules for the memory system, and the identity relation for store buffer steps, the program steps are still a parameter. 
We refer to a transition by \isa{{\isacharparenleft}\freeify{ts}{\isacharcomma}\ \freeify{{\isasymS}}{\isacharcomma}\ \freeify{m}{\isacharparenright}\ $\overset{\isa{v}}{\Rightarrow}$\ {\isacharparenleft}\freeify{ts{\isacharprime}}{\isacharcomma}\ \freeify{{\isasymS}{\isacharprime}}{\isacharcomma}\ \freeify{m{\isacharprime}}{\isacharparenright}}.
\begin{figure}
\begin{center}
\isa{\mbox{}\inferrule{\mbox{}}{\mbox{{\isacharparenleft}\constructor{Read}\ \freeify{volatile}\ \freeify{a}\ \freeify{t}\ {\isasymcdot}\ \freeify{is}{\isacharcomma}\ \freeify{{\isasymtheta}}{\isacharcomma}\ \freeify{x}{\isacharcomma}\ \freeify{m}{\isacharcomma}\ \freeify{{\isasymG}}{\isacharparenright}\ $\overset{\isa{v}}{\rightarrow}_{\isa{m}}$\ {\isacharparenleft}\freeify{is}{\isacharcomma}\ \freeify{{\isasymtheta}}{\isacharparenleft}\freeify{t}\ {\isasymmapsto}\ \freeify{m}\ \freeify{a}{\isacharparenright}{\isacharcomma}\ \freeify{x}{\isacharcomma}\ \freeify{m}{\isacharcomma}\ \freeify{{\isasymG}}{\isacharparenright}}}}\\[-0.3\baselineskip]
\isa{\mbox{}\inferrule{\mbox{}}{\mbox{{\isacharparenleft}\constructor{Write}\ False\ \freeify{a}\ {\isacharparenleft}\freeify{D}{\isacharcomma}\ \freeify{f}{\isacharparenright}\ \freeify{A}\ \freeify{L}\ \freeify{R}\ \freeify{W}\ {\isasymcdot}\ \freeify{is}{\isacharcomma}\ \freeify{{\isasymtheta}}{\isacharcomma}\ \freeify{x}{\isacharcomma}\ \freeify{m}{\isacharcomma}\ \freeify{{\isasymG}}{\isacharparenright}\ $\overset{\isa{v}}{\rightarrow}_{\isa{m}}$\ {\isacharparenleft}\freeify{is}{\isacharcomma}\ \freeify{{\isasymtheta}}{\isacharcomma}\ \freeify{x}{\isacharcomma}\ \freeify{m}{\isacharparenleft}\freeify{a}\ {\isacharcolon}{\isacharequal}\ \freeify{f}\ \freeify{{\isasymtheta}}{\isacharparenright}{\isacharcomma}\ \freeify{{\isasymG}}{\isacharparenright}}}}\\[0.5\baselineskip]
\isa{\mbox{}\inferrule{\mbox{\freeify{{\isasymG}}\ {\isacharequal}\ {\isacharparenleft}\freeify{{\isasymD}}{\isacharcomma}\ \freeify{{\isasymO}}{\isacharcomma}\ \freeify{{\isasymA}}{\isacharcomma}\ \freeify{{\isasymS}}{\isacharparenright}}\\\ \mbox{\freeify{{\isasymG}{\isacharprime}}\ {\isacharequal}\ {\isacharparenleft}True{\isacharcomma}\ \freeify{{\isasymO}}\ {\isasymunion}\ \freeify{A}\ {\isacharminus}\ \freeify{R}{\isacharcomma}\ \freeify{{\isasymA}}\ {\isasymunion}\ \freeify{A}\ {\isacharminus}\ \freeify{R}{\isacharcomma}\ \freeify{{\isasymS}}\ {\isasymoplus}\isactrlbsub \freeify{W}\isactrlesub \ \freeify{R}\ {\isasymominus}\isactrlbsub \freeify{A}\isactrlesub \ \freeify{L}{\isacharparenright}}}{\mbox{{\isacharparenleft}\constructor{Write}\ True\ \freeify{a}\ {\isacharparenleft}\freeify{D}{\isacharcomma}\ \freeify{f}{\isacharparenright}\ \freeify{A}\ \freeify{L}\ \freeify{R}\ \freeify{W}\ {\isasymcdot}\ \freeify{is}{\isacharcomma}\ \freeify{{\isasymtheta}}{\isacharcomma}\ \freeify{x}{\isacharcomma}\ \freeify{m}{\isacharcomma}\ \freeify{{\isasymG}}{\isacharparenright}\ $\overset{\isa{v}}{\rightarrow}_{\isa{m}}$\ {\isacharparenleft}\freeify{is}{\isacharcomma}\ \freeify{{\isasymtheta}}{\isacharcomma}\ \freeify{x}{\isacharcomma}\ \freeify{m}{\isacharparenleft}\freeify{a}\ {\isacharcolon}{\isacharequal}\ \freeify{f}\ \freeify{{\isasymtheta}}{\isacharparenright}{\isacharcomma}\ \freeify{{\isasymG}{\isacharprime}}{\isacharparenright}}}}\\[0.5\baselineskip]
\isa{\mbox{}\inferrule{\mbox{{\isasymnot}\ \freeify{cond}\ {\isacharparenleft}\freeify{{\isasymtheta}}{\isacharparenleft}\freeify{t}\ {\isasymmapsto}\ \freeify{m}\ \freeify{a}{\isacharparenright}{\isacharparenright}}\\\ \mbox{\freeify{{\isasymG}}\ {\isacharequal}\ {\isacharparenleft}\freeify{{\isasymD}}{\isacharcomma}\ \freeify{{\isasymO}}{\isacharcomma}\ \freeify{{\isasymA}}{\isacharcomma}\ \freeify{{\isasymS}}{\isacharparenright}}\\\ \mbox{\freeify{{\isasymG}{\isacharprime}}\ {\isacharequal}\ {\isacharparenleft}False{\isacharcomma}\ \freeify{{\isasymO}}{\isacharcomma}\ {\isasymemptyset}{\isacharcomma}\ \freeify{{\isasymS}}{\isacharparenright}}}{\mbox{{\isacharparenleft}\constructor{RMW}\ \freeify{a}\ \freeify{t}\ {\isacharparenleft}\freeify{D}{\isacharcomma}\ \freeify{f}{\isacharparenright}\ \freeify{cond}\ \freeify{ret}\ \freeify{A}\ \freeify{L}\ \freeify{R}\ \freeify{W}\ {\isasymcdot}\ \freeify{is}{\isacharcomma}\ \freeify{{\isasymtheta}}{\isacharcomma}\ \freeify{x}{\isacharcomma}\ \freeify{m}{\isacharcomma}\ \freeify{{\isasymG}}{\isacharparenright}\ $\overset{\isa{v}}{\rightarrow}_{\isa{m}}$\ {\isacharparenleft}\freeify{is}{\isacharcomma}\ \freeify{{\isasymtheta}}{\isacharparenleft}\freeify{t}\ {\isasymmapsto}\ \freeify{m}\ \freeify{a}{\isacharparenright}{\isacharcomma}\ \freeify{x}{\isacharcomma}\ \freeify{m}{\isacharcomma}\ \freeify{{\isasymG}{\isacharprime}}{\isacharparenright}}}}\\[0.5\baselineskip]
\isa{\mbox{}\inferrule{\mbox{\freeify{cond}\ {\isacharparenleft}\freeify{{\isasymtheta}}{\isacharparenleft}\freeify{t}\ {\isasymmapsto}\ \freeify{m}\ \freeify{a}{\isacharparenright}{\isacharparenright}}\\\ \mbox{\freeify{{\isasymtheta}{\isacharprime}}\ {\isacharequal}\ \freeify{{\isasymtheta}}{\isacharparenleft}\freeify{t}\ {\isasymmapsto}\ \freeify{ret}\ {\isacharparenleft}\freeify{m}\ \freeify{a}{\isacharparenright}\ {\isacharparenleft}\freeify{f}\ {\isacharparenleft}\freeify{{\isasymtheta}}{\isacharparenleft}\freeify{t}\ {\isasymmapsto}\ \freeify{m}\ \freeify{a}{\isacharparenright}{\isacharparenright}{\isacharparenright}{\isacharparenright}}\\\ \mbox{\freeify{m{\isacharprime}}\ {\isacharequal}\ \freeify{m}{\isacharparenleft}\freeify{a}\ {\isacharcolon}{\isacharequal}\ \freeify{f}\ {\isacharparenleft}\freeify{{\isasymtheta}}{\isacharparenleft}\freeify{t}\ {\isasymmapsto}\ \freeify{m}\ \freeify{a}{\isacharparenright}{\isacharparenright}{\isacharparenright}}\\\ \mbox{\freeify{{\isasymG}}\ {\isacharequal}\ {\isacharparenleft}\freeify{{\isasymD}}{\isacharcomma}\ \freeify{{\isasymO}}{\isacharcomma}\ \freeify{{\isasymA}}{\isacharcomma}\ \freeify{{\isasymS}}{\isacharparenright}}\\\ \mbox{\freeify{{\isasymG}{\isacharprime}}\ {\isacharequal}\ {\isacharparenleft}False{\isacharcomma}\ \freeify{{\isasymO}}\ {\isasymunion}\ \freeify{A}\ {\isacharminus}\ \freeify{R}{\isacharcomma}\ {\isasymemptyset}{\isacharcomma}\ \freeify{{\isasymS}}\ {\isasymoplus}\isactrlbsub \freeify{W}\isactrlesub \ \freeify{R}\ {\isasymominus}\isactrlbsub \freeify{A}\isactrlesub \ \freeify{L}{\isacharparenright}}}{\mbox{{\isacharparenleft}\constructor{RMW}\ \freeify{a}\ \freeify{t}\ {\isacharparenleft}\freeify{D}{\isacharcomma}\ \freeify{f}{\isacharparenright}\ \freeify{cond}\ \freeify{ret}\ \freeify{A}\ \freeify{L}\ \freeify{R}\ \freeify{W}\ {\isasymcdot}\ \freeify{is}{\isacharcomma}\ \freeify{{\isasymtheta}}{\isacharcomma}\ \freeify{x}{\isacharcomma}\ \freeify{m}{\isacharcomma}\ \freeify{{\isasymG}}{\isacharparenright}\ $\overset{\isa{v}}{\rightarrow}_{\isa{m}}$\ {\isacharparenleft}\freeify{is}{\isacharcomma}\ \freeify{{\isasymtheta}{\isacharprime}}{\isacharcomma}\ \freeify{x}{\isacharcomma}\ \freeify{m{\isacharprime}}{\isacharcomma}\ \freeify{{\isasymG}{\isacharprime}}{\isacharparenright}}}}\\[-0.3\baselineskip]
\isa{\mbox{}\inferrule{\mbox{}}{\mbox{{\isacharparenleft}\constructor{Fence}\ {\isasymcdot}\ \freeify{is}{\isacharcomma}\ \freeify{{\isasymtheta}}{\isacharcomma}\ \freeify{x}{\isacharcomma}\ \freeify{m}{\isacharcomma}\ \freeify{{\isasymD}}{\isacharcomma}\ \freeify{{\isasymO}}{\isacharcomma}\ \freeify{{\isasymA}}{\isacharcomma}\ \freeify{{\isasymS}}{\isacharparenright}\ $\overset{\isa{v}}{\rightarrow}_{\isa{m}}$\ {\isacharparenleft}\freeify{is}{\isacharcomma}\ \freeify{{\isasymtheta}}{\isacharcomma}\ \freeify{x}{\isacharcomma}\ \freeify{m}{\isacharcomma}\ False{\isacharcomma}\ \freeify{{\isasymO}}{\isacharcomma}\ {\isasymemptyset}{\isacharcomma}\ \freeify{{\isasymS}}{\isacharparenright}}}}\\[-0.3\baselineskip]
\isa{\mbox{}\inferrule{\mbox{}}{\mbox{{\isacharparenleft}\constructor{Ghost}\ \freeify{A}\ \freeify{L}\ {\isasymcdot}\ \freeify{is}{\isacharcomma}\ \freeify{{\isasymtheta}}{\isacharcomma}\ \freeify{x}{\isacharcomma}\ \freeify{m}{\isacharcomma}\ \freeify{{\isasymD}}{\isacharcomma}\ \freeify{{\isasymO}}{\isacharcomma}\ \freeify{{\isasymA}}{\isacharcomma}\ \freeify{{\isasymS}}{\isacharparenright}\ $\overset{\isa{v}}{\rightarrow}_{\isa{m}}$\ {\isacharparenleft}\freeify{is}{\isacharcomma}\ \freeify{{\isasymtheta}}{\isacharcomma}\ \freeify{x}{\isacharcomma}\ \freeify{m}{\isacharcomma}\ \freeify{{\isasymD}}{\isacharcomma}\ \freeify{{\isasymO}}\ {\isasymunion}\ \freeify{A}{\isacharcomma}\ \freeify{{\isasymA}}\ {\isasymunion}\ \freeify{A}{\isacharcomma}\ \freeify{{\isasymS}}\ {\isasymominus}\isactrlbsub \freeify{A}\isactrlesub \ \freeify{L}{\isacharparenright}}}}\\[0.1\baselineskip]

\end{center}
\caption{Memory transitions of the virtual machine \label{fig:direct-memory}}
\end{figure}

In addition to the transition rules for the virtual machine we introduce the \emph{safety} judgment \isa{\freeify{{\isasymO}s}{\isacharcomma}\freeify{i}{\isasymturnstile}\ {\isacharparenleft}\freeify{is}{\isacharcomma}\ \freeify{{\isasymtheta}}{\isacharcomma}\ \freeify{x}{\isacharcomma}\ \freeify{m}{\isacharcomma}\ \freeify{{\isasymD}}{\isacharcomma}\ \freeify{{\isasymO}}{\isacharcomma}\ \freeify{{\isasymA}}{\isacharcomma}\ \freeify{{\isasymS}}{\isacharparenright}\ {\isasymsurd}} in Figure~\ref{fig:safe-direct-memory}, where \isa{\freeify{{\isasymO}s}} is the list of ownership sets obtained from the thread list \isa{\freeify{ts}} and \isa{\freeify{i}} is the thread index.
Safety of all reachable states of the virtual machine ensures that the access policy is obeyed by the program and is our formal 
prerequisite for the simulation theorem.
It is left as a proof obligation to be discharged by means of a proper program logic for sequentially consistent executions.
\begin{figure}
\begin{center}
\isa{\mbox{}\inferrule{\mbox{\freeify{a}\ {\isasymin}\ \freeify{{\isasymO}}\ {\isasymor}\ \freeify{a}\ {\isasymin}\ read{\isacharunderscore}only\ \freeify{{\isasymS}}\ {\isasymor}\ \freeify{volatile}\ {\isasymand}\ \freeify{a}\ {\isasymin}\ dom\ \freeify{{\isasymS}}}\\\ \mbox{\freeify{volatile}\ {\isasymlongrightarrow}\ {\isasymnot}\ \freeify{{\isasymD}}}\\\ \mbox{{\isasymnot}\ \freeify{volatile}\ {\isasymlongrightarrow}\ \freeify{a}\ {\isasymin}\ \freeify{{\isasymA}}\ {\isasymlongrightarrow}\ {\isasymnot}\ \freeify{{\isasymD}}}}{\mbox{\freeify{{\isasymO}s}{\isacharcomma}\freeify{i}{\isasymturnstile}\ {\isacharparenleft}\constructor{Read}\ \freeify{volatile}\ \freeify{a}\ \freeify{t}\ {\isasymcdot}\ \freeify{is}{\isacharcomma}\ \freeify{{\isasymtheta}}{\isacharcomma}\ \freeify{x}{\isacharcomma}\ \freeify{m}{\isacharcomma}\ \freeify{{\isasymD}}{\isacharcomma}\ \freeify{{\isasymO}}{\isacharcomma}\ \freeify{{\isasymA}}{\isacharcomma}\ \freeify{{\isasymS}}{\isacharparenright}\ {\isasymsurd}}}}\\[0.5\baselineskip]
\isa{\mbox{}\inferrule{\mbox{\freeify{a}\ {\isasymin}\ \freeify{{\isasymO}}}\\\ \mbox{\freeify{a}\ {\isasymnotin}\ dom\ \freeify{{\isasymS}}}}{\mbox{\freeify{{\isasymO}s}{\isacharcomma}\freeify{i}{\isasymturnstile}\ {\isacharparenleft}\constructor{Write}\ False\ \freeify{a}\ {\isacharparenleft}\freeify{D}{\isacharcomma}\ \freeify{f}{\isacharparenright}\ \freeify{A}\ \freeify{L}\ \freeify{R}\ \freeify{W}\ {\isasymcdot}\ \freeify{is}{\isacharcomma}\ \freeify{{\isasymtheta}}{\isacharcomma}\ \freeify{x}{\isacharcomma}\ \freeify{m}{\isacharcomma}\ \freeify{{\isasymD}}{\isacharcomma}\ \freeify{{\isasymO}}{\isacharcomma}\ \freeify{{\isasymA}}{\isacharcomma}\ \freeify{{\isasymS}}{\isacharparenright}\ {\isasymsurd}}}}\\[0.5\baselineskip]
\isa{\mbox{}\inferrule{\mbox{{\isasymforall}\boundify{j}{\isacharless}{\isacharbar}\freeify{{\isasymO}s}{\isacharbar}{\isachardot}\ \freeify{i}\ {\isasymnoteq}\ \boundify{j}\ {\isasymlongrightarrow}\ \freeify{a}\ {\isasymnotin}\ \freeify{{\isasymO}s}\ensuremath{_{[\boundify{j}]}}}\\\ \mbox{\freeify{a}\ {\isasymnotin}\ read{\isacharunderscore}only\ \freeify{{\isasymS}}}\\\ \mbox{{\isasymforall}\boundify{j}{\isacharless}{\isacharbar}\freeify{{\isasymO}s}{\isacharbar}{\isachardot}\ \freeify{i}\ {\isasymnoteq}\ \boundify{j}\ {\isasymlongrightarrow}\ \freeify{A}\ {\isasyminter}\ \freeify{{\isasymO}s}\ensuremath{_{[\boundify{j}]}}\ {\isacharequal}\ {\isasymemptyset}}\\\ \mbox{\freeify{A}\ {\isasymsubseteq}\ \freeify{{\isasymO}}\ {\isasymunion}\ dom\ \freeify{{\isasymS}}}\\\ \mbox{\freeify{L}\ {\isasymsubseteq}\ \freeify{A}}\\\ \mbox{\freeify{R}\ {\isasymsubseteq}\ \freeify{{\isasymO}}}\\\ \mbox{\freeify{A}\ {\isasyminter}\ \freeify{R}\ {\isacharequal}\ {\isasymemptyset}}}{\mbox{\freeify{{\isasymO}s}{\isacharcomma}\freeify{i}{\isasymturnstile}\ {\isacharparenleft}\constructor{Write}\ True\ \freeify{a}\ {\isacharparenleft}\freeify{D}{\isacharcomma}\ \freeify{f}{\isacharparenright}\ \freeify{A}\ \freeify{L}\ \freeify{R}\ \freeify{W}\ {\isasymcdot}\ \freeify{is}{\isacharcomma}\ \freeify{{\isasymtheta}}{\isacharcomma}\ \freeify{x}{\isacharcomma}\ \freeify{m}{\isacharcomma}\ \freeify{{\isasymD}}{\isacharcomma}\ \freeify{{\isasymO}}{\isacharcomma}\ \freeify{{\isasymA}}{\isacharcomma}\ \freeify{{\isasymS}}{\isacharparenright}\ {\isasymsurd}}}}\\[0.5\baselineskip]
\isa{\mbox{}\inferrule{\mbox{{\isasymnot}\ \freeify{cond}\ {\isacharparenleft}\freeify{{\isasymtheta}}{\isacharparenleft}\freeify{t}\ {\isasymmapsto}\ \freeify{m}\ \freeify{a}{\isacharparenright}{\isacharparenright}}\\\ \mbox{\freeify{a}\ {\isasymin}\ dom\ \freeify{{\isasymS}}\ {\isasymunion}\ \freeify{{\isasymO}}}}{\mbox{\freeify{{\isasymO}s}{\isacharcomma}\freeify{i}{\isasymturnstile}\ {\isacharparenleft}\constructor{RMW}\ \freeify{a}\ \freeify{t}\ {\isacharparenleft}\freeify{D}{\isacharcomma}\ \freeify{f}{\isacharparenright}\ \freeify{cond}\ \freeify{ret}\ \freeify{A}\ \freeify{L}\ \freeify{R}\ \freeify{W}\ {\isasymcdot}\ \freeify{is}{\isacharcomma}\ \freeify{{\isasymtheta}}{\isacharcomma}\ \freeify{x}{\isacharcomma}\ \freeify{m}{\isacharcomma}\ \freeify{{\isasymD}}{\isacharcomma}\ \freeify{{\isasymO}}{\isacharcomma}\ \freeify{{\isasymA}}{\isacharcomma}\ \freeify{{\isasymS}}{\isacharparenright}\ {\isasymsurd}}}}\\[0.5\baselineskip]
\isa{\mbox{}\inferrule{\mbox{\freeify{cond}\ {\isacharparenleft}\freeify{{\isasymtheta}}{\isacharparenleft}\freeify{t}\ {\isasymmapsto}\ \freeify{m}\ \freeify{a}{\isacharparenright}{\isacharparenright}}\\\ \mbox{{\isasymforall}\boundify{j}{\isacharless}{\isacharbar}\freeify{{\isasymO}s}{\isacharbar}{\isachardot}\ \freeify{i}\ {\isasymnoteq}\ \boundify{j}\ {\isasymlongrightarrow}\ \freeify{a}\ {\isasymnotin}\ \freeify{{\isasymO}s}\ensuremath{_{[\boundify{j}]}}}\\\ \mbox{\freeify{a}\ {\isasymnotin}\ read{\isacharunderscore}only\ \freeify{{\isasymS}}}\\\ \mbox{{\isasymforall}\boundify{j}{\isacharless}{\isacharbar}\freeify{{\isasymO}s}{\isacharbar}{\isachardot}\ \freeify{i}\ {\isasymnoteq}\ \boundify{j}\ {\isasymlongrightarrow}\ \freeify{A}\ {\isasyminter}\ \freeify{{\isasymO}s}\ensuremath{_{[\boundify{j}]}}\ {\isacharequal}\ {\isasymemptyset}}\\\ \mbox{\freeify{A}\ {\isasymsubseteq}\ \freeify{{\isasymO}}\ {\isasymunion}\ dom\ \freeify{{\isasymS}}}\\\ \mbox{\freeify{L}\ {\isasymsubseteq}\ \freeify{A}}\\\ \mbox{\freeify{R}\ {\isasymsubseteq}\ \freeify{{\isasymO}}}\\\ \mbox{\freeify{A}\ {\isasyminter}\ \freeify{R}\ {\isacharequal}\ {\isasymemptyset}}}{\mbox{\freeify{{\isasymO}s}{\isacharcomma}\freeify{i}{\isasymturnstile}\ {\isacharparenleft}\constructor{RMW}\ \freeify{a}\ \freeify{t}\ {\isacharparenleft}\freeify{D}{\isacharcomma}\ \freeify{f}{\isacharparenright}\ \freeify{cond}\ \freeify{ret}\ \freeify{A}\ \freeify{L}\ \freeify{R}\ \freeify{W}\ {\isasymcdot}\ \freeify{is}{\isacharcomma}\ \freeify{{\isasymtheta}}{\isacharcomma}\ \freeify{x}{\isacharcomma}\ \freeify{m}{\isacharcomma}\ \freeify{{\isasymD}}{\isacharcomma}\ \freeify{{\isasymO}}{\isacharcomma}\ \freeify{{\isasymA}}{\isacharcomma}\ \freeify{{\isasymS}}{\isacharparenright}\ {\isasymsurd}}}}\\[-0.3\baselineskip]
\isa{\mbox{}\inferrule{\mbox{}}{\mbox{\freeify{{\isasymO}s}{\isacharcomma}\freeify{i}{\isasymturnstile}\ {\isacharparenleft}\constructor{Fence}\ {\isasymcdot}\ \freeify{is}{\isacharcomma}\ \freeify{{\isasymtheta}}{\isacharcomma}\ \freeify{x}{\isacharcomma}\ \freeify{m}{\isacharcomma}\ \freeify{{\isasymD}}{\isacharcomma}\ \freeify{{\isasymO}}{\isacharcomma}\ \freeify{{\isasymA}}{\isacharcomma}\ \freeify{{\isasymS}}{\isacharparenright}\ {\isasymsurd}}}}\\[0.5\baselineskip]
\isa{\mbox{}\inferrule{\mbox{\freeify{A}\ {\isasymsubseteq}\ dom\ \freeify{{\isasymS}}\ {\isasymunion}\ \freeify{{\isasymO}}}\\\ \mbox{\freeify{L}\ {\isasymsubseteq}\ \freeify{A}}\\\ \mbox{{\isasymforall}\boundify{j}{\isacharless}{\isacharbar}\freeify{{\isasymO}s}{\isacharbar}{\isachardot}\ \freeify{i}\ {\isasymnoteq}\ \boundify{j}\ {\isasymlongrightarrow}\ \freeify{A}\ {\isasyminter}\ \freeify{{\isasymO}s}\ensuremath{_{[\boundify{j}]}}\ {\isacharequal}\ {\isasymemptyset}}}{\mbox{\freeify{{\isasymO}s}{\isacharcomma}\freeify{i}{\isasymturnstile}\ {\isacharparenleft}\constructor{Ghost}\ \freeify{A}\ \freeify{L}\ {\isasymcdot}\ \freeify{is}{\isacharcomma}\ \freeify{{\isasymtheta}}{\isacharcomma}\ \freeify{x}{\isacharcomma}\ \freeify{m}{\isacharcomma}\ \freeify{{\isasymD}}{\isacharcomma}\ \freeify{{\isasymO}}{\isacharcomma}\ \freeify{{\isasymA}}{\isacharcomma}\ \freeify{{\isasymS}}{\isacharparenright}\ {\isasymsurd}}}}\\[0.1\baselineskip]
\end{center}
\caption{Safe configurations of a virtual machine \label{fig:safe-direct-memory}}
\end{figure}
In the following we elaborate on the rules of Figures \ref{fig:direct-memory} and \ref{fig:safe-direct-memory} in parallel.
To read from an address it either has to be owned or read-only or it has to be volatile and shared. Moreover the read has to be clean.
The memory content of address \isa{\freeify{a}} is stored in temporary \isa{\freeify{t}}. 
Non-volatile writes are only allowed to owned and unshared addresses. 
The result is written directly into the memory. 
A volatile write is only allowed when no other thread owns the address and the address is not marked as read-only.
Simultaneously with the volatile write we can transfer ownership as specified by the annotations \isa{\freeify{A}}, \isa{\freeify{L}}, \isa{\freeify{R}} and \isa{\freeify{W}}. 
The acquired addresses \isa{\freeify{A}} must not be owned by any other thread and stem from the shared addresses or are already owned.
Reacquiring owned addresses can be used to change the shared-status via the set of local addresses \isa{\freeify{L}} which have to be a subset of \isa{\freeify{A}}. 
The released addresses \isa{\freeify{R}} have to be owned and distinct from the acquired addresses \isa{\freeify{A}}. 
After the write the new ownership of the thread is obtained by adding the acquired addresses \isa{\freeify{A}} and releasing the addresses \isa{\freeify{R}}: \isa{\freeify{{\isasymO}}\ {\isasymunion}\ \freeify{A}\ {\isacharminus}\ \freeify{R}}. Analogously the set of acquired addresses \isa{\freeify{{\isasymA}}} is updated. The released addresses \isa{\freeify{R}} are augmented to the shared addresses \isa{\freeify{S}} and the local addresses \isa{\freeify{L}} are removed. We also take care about the write permissions in the shared state: the released addresses in set \isa{\freeify{W}} as well as the acquired addresses are marked writable: \isa{\freeify{{\isasymS}}\ {\isasymoplus}\isactrlbsub \freeify{W}\isactrlesub \ \freeify{R}\ {\isasymominus}\isactrlbsub \freeify{A}\isactrlesub \ \freeify{L}}. The auxiliary ternary operators to augment and subtract addresses from the sharing map are defined as follows:

\begin{isabelle}%
\freeify{{\isasymS}}\ {\isasymoplus}\isactrlbsub \freeify{W}\isactrlesub \ \freeify{R}\ {\isasymequiv}\ {\isasymlambda}\boundify{a}{\isachardot}\ \holkeyword{if}\ \boundify{a}\ {\isasymin}\ \freeify{R}\ \holkeyword{then}\ {\isasymlfloor}\boundify{a}\ {\isasymin}\ \freeify{W}{\isasymrfloor}\ \holkeyword{else}\ \freeify{{\isasymS}}\ \boundify{a}%
\end{isabelle}
\begin{isabelle}%
\freeify{{\isasymS}}\ {\isasymominus}\isactrlbsub \freeify{A}\isactrlesub \ \freeify{L}\ {\isasymequiv}\isanewline
\isaindent{\ \ }{\isasymlambda}\boundify{a}{\isachardot}\ \holkeyword{if}\ \boundify{a}\ {\isasymin}\ \freeify{L}\ \holkeyword{then}\ {\isasymbottom}\ \holkeyword{else}\ \holkeyword{case}\ \freeify{{\isasymS}}\ \boundify{a}\ \holkeyword{of}\ {\isasymbottom}\ {\isasymRightarrow}\ {\isasymbottom}\ {\isacharbar}\ {\isasymlfloor}\boundify{writeable}{\isasymrfloor}\ {\isasymRightarrow}\ {\isasymlfloor}\boundify{a}\ {\isasymin}\ \freeify{A}\ {\isasymor}\ \boundify{writeable}{\isasymrfloor}%
\end{isabelle}

The read-modify-write instruction first adds the current value at address \isa{\freeify{a}} to temporary \isa{\freeify{t}} and then checks the store condition \isa{\freeify{cond}} on the temporaries. 
If it fails this read is the final result of the operation. 
Otherwise the store is performed. 
The resulting value of the temporary \isa{\freeify{t}} is specified by the function \isa{\freeify{ret}} which considers both the old and new value as input. 
As the read-modify-write instruction is an interlocked operation which flushes the store buffer as a side effect the dirty flag \isa{\freeify{{\isasymD}}} as well as the set of acquired addresses \isa{\freeify{{\isasymA}}} are reset.
The other effects on the ghost state and the safety sideconditions are the same as for the volatile read and volatile write, respectively.

The only effect of the fence instruction in the system without store buffer is to reset the dirty flag and the set of acquired addresses.

The ghost instruction \isa{\constructor{Ghost}\ \freeify{A}\ \freeify{L}} allows to acquire ownership when no write is involved \ie when merely reading from memory. It has the same safety requirements as the corresponding parts in the write instructions. Releasing ownership can always be delayed to the next volatile (or interlocked) write instruction, since only with the write another thread can gain information about released addresses. In the simulation proof we build on the fact that all pending ghost operations in the store buffer until the first volatile write may only acquire addresses to the ownership of the thread.%
\end{isamarkuptext}%
\isamarkuptrue%
\isamarkupsubsection{Store buffer machine \label{sec:storebuffermachine}%
}
\isamarkuptrue%
\begin{isamarkuptext}%
The store buffer machine extends the virtual machine by maintaining a list of outstanding memory writes.
Write instructions are appended to the store buffer and read instructions are satisfied from the store buffer if possible. 
To support our coupling relation between a configuration of the store buffer machine and a configuration of the virtual machine, we also maintain additional bookkeeping information inside the store buffer. 
For every write we keep the volatile flag and the store operation. Moreover we record read,  program and ghost steps.
This allows us to restore the necessary computation history of the store buffer machine and relate it to the virtual machine which may fall behind the store buffer machine during execution. 
Altogether an entry in the store buffer is either a
\begin{itemize}
\item \isa{\constructor{Read\isactrlisub s\isactrlisub b}\ \freeify{volatile}\ \freeify{a}\ \freeify{t}\ \freeify{v}}, recording a corresponding read from address \isa{\freeify{a}} which loaded the value \isa{\freeify{v}} to temporary \isa{\freeify{t}}, or a 
\item \isa{\constructor{Write\isactrlisub s\isactrlisub b}\ \freeify{volatile}\ \freeify{a}\ \freeify{sop}\ \freeify{v}} for an outstanding write, where operation \isa{\freeify{sop}} evaluated to value \isa{\freeify{v}}, or of the form

\item \isa{\constructor{Prog\isactrlisub s\isactrlisub b}\ \freeify{p}\ \freeify{p{\isacharprime}}\ \freeify{is{\isacharprime}}}, recording a program transition from \isa{\freeify{p}} to \isa{\freeify{p{\isacharprime}}} which issued instructions \isa{\freeify{is{\isacharprime}}}, or of the form
\item \isa{\constructor{Ghost\isactrlisub s\isactrlisub b}\ \freeify{A}\ \freeify{L}}, recording a corresponding ghost operation to acquire addresses \isa{\freeify{A}} and keep addresses \isa{\freeify{L}} local.
\end{itemize}
As defined in Figure \ref{fig:store-buffer-transitions} a write updates the memory when it exits the store buffer, all other store buffer entries may only have an effect on the ghost state. The effect on the ownership information is analogous to the corresponding operations in the virtual machine.
\begin{figure}
\begin{center}
\isa{\mbox{}\inferrule{\mbox{}}{\mbox{{\isacharparenleft}\freeify{m}{\isacharcomma}\ \constructor{Write\isactrlisub s\isactrlisub b}\ False\ \freeify{a}\ \freeify{sop}\ \freeify{v}\ \freeify{A}\ \freeify{L}\ \freeify{R}\ \freeify{W}\ {\isasymcdot}\ \freeify{sb}{\isacharcomma}\ \freeify{{\isasymO}}{\isacharcomma}\ \freeify{{\isasymA}}{\isacharcomma}\ \freeify{{\isasymS}}{\isacharparenright}\ {\isasymrightarrow}\isactrlsub s\isactrlsub b\ {\isacharparenleft}\freeify{m}{\isacharparenleft}\freeify{a}\ {\isacharcolon}{\isacharequal}\ \freeify{v}{\isacharparenright}{\isacharcomma}\ \freeify{sb}{\isacharcomma}\ \freeify{{\isasymO}}{\isacharcomma}\ \freeify{{\isasymA}}{\isacharcomma}\ \freeify{{\isasymS}}{\isacharparenright}}}}\\[0.5\baselineskip]
\isa{\mbox{}\inferrule{\mbox{\freeify{{\isasymO}{\isacharprime}}\ {\isacharequal}\ \freeify{{\isasymO}}\ {\isasymunion}\ \freeify{A}\ {\isacharminus}\ \freeify{R}}\\\ \mbox{\freeify{{\isasymA}{\isacharprime}}\ {\isacharequal}\ \freeify{{\isasymA}}\ {\isasymunion}\ \freeify{A}\ {\isacharminus}\ \freeify{R}}\\\ \mbox{\freeify{{\isasymS}{\isacharprime}}\ {\isacharequal}\ \freeify{{\isasymS}}\ {\isasymoplus}\isactrlbsub \freeify{W}\isactrlesub \ \freeify{R}\ {\isasymominus}\isactrlbsub \freeify{A}\isactrlesub \ \freeify{L}}}{\mbox{{\isacharparenleft}\freeify{m}{\isacharcomma}\ \constructor{Write\isactrlisub s\isactrlisub b}\ True\ \freeify{a}\ \freeify{sop}\ \freeify{v}\ \freeify{A}\ \freeify{L}\ \freeify{R}\ \freeify{W}\ {\isasymcdot}\ \freeify{sb}{\isacharcomma}\ \freeify{{\isasymO}}{\isacharcomma}\ \freeify{{\isasymA}}{\isacharcomma}\ \freeify{{\isasymS}}{\isacharparenright}\ {\isasymrightarrow}\isactrlsub s\isactrlsub b\ {\isacharparenleft}\freeify{m}{\isacharparenleft}\freeify{a}\ {\isacharcolon}{\isacharequal}\ \freeify{v}{\isacharparenright}{\isacharcomma}\ \freeify{sb}{\isacharcomma}\ \freeify{{\isasymO}{\isacharprime}}{\isacharcomma}\ \freeify{{\isasymA}{\isacharprime}}{\isacharcomma}\ \freeify{{\isasymS}{\isacharprime}}{\isacharparenright}}}}\\[-0.3\baselineskip]
\isa{\mbox{}\inferrule{\mbox{}}{\mbox{{\isacharparenleft}\freeify{m}{\isacharcomma}\ \constructor{Read\isactrlisub s\isactrlisub b}\ \freeify{volatile}\ \freeify{a}\ \freeify{t}\ \freeify{v}\ {\isasymcdot}\ \freeify{sb}{\isacharcomma}\ \freeify{{\isasymO}}{\isacharcomma}\ \freeify{{\isasymA}}{\isacharcomma}\ \freeify{{\isasymS}}{\isacharparenright}\ {\isasymrightarrow}\isactrlsub s\isactrlsub b\ {\isacharparenleft}\freeify{m}{\isacharcomma}\ \freeify{sb}{\isacharcomma}\ \freeify{{\isasymO}}{\isacharcomma}\ \freeify{{\isasymA}}{\isacharcomma}\ \freeify{{\isasymS}}{\isacharparenright}}}}\\[-0.3\baselineskip]
\isa{\mbox{}\inferrule{\mbox{}}{\mbox{{\isacharparenleft}\freeify{m}{\isacharcomma}\ \constructor{Prog\isactrlisub s\isactrlisub b}\ \freeify{p}\ \freeify{p{\isacharprime}}\ \freeify{is}\ {\isasymcdot}\ \freeify{sb}{\isacharcomma}\ \freeify{{\isasymO}}{\isacharcomma}\ \freeify{{\isasymA}}{\isacharcomma}\ \freeify{{\isasymS}}{\isacharparenright}\ {\isasymrightarrow}\isactrlsub s\isactrlsub b\ {\isacharparenleft}\freeify{m}{\isacharcomma}\ \freeify{sb}{\isacharcomma}\ \freeify{{\isasymO}}{\isacharcomma}\ \freeify{{\isasymA}}{\isacharcomma}\ \freeify{{\isasymS}}{\isacharparenright}}}}\\[-0.3\baselineskip]
\isa{\mbox{}\inferrule{\mbox{}}{\mbox{{\isacharparenleft}\freeify{m}{\isacharcomma}\ \constructor{Ghost\isactrlisub s\isactrlisub b}\ \freeify{A}\ \freeify{L}\ {\isasymcdot}\ \freeify{sb}{\isacharcomma}\ \freeify{{\isasymO}}{\isacharcomma}\ \freeify{{\isasymA}}{\isacharcomma}\ \freeify{{\isasymS}}{\isacharparenright}\ {\isasymrightarrow}\isactrlsub s\isactrlsub b\ {\isacharparenleft}\freeify{m}{\isacharcomma}\ \freeify{sb}{\isacharcomma}\ \freeify{{\isasymO}}\ {\isasymunion}\ \freeify{A}{\isacharcomma}\ \freeify{{\isasymA}}\ {\isasymunion}\ \freeify{A}{\isacharcomma}\ \freeify{{\isasymS}}\ {\isasymominus}\isactrlbsub \freeify{A}\isactrlesub \ \freeify{L}{\isacharparenright}}}} 
\end{center}
\caption{Store buffer transitions \label{fig:store-buffer-transitions}}
\end{figure}
The transitions defined in Figure \ref{fig:store-buffer-memory} are straightforward extensions of the transitions of the virtual machine. 
With \isa{buffered{\isacharunderscore}val\ \freeify{sb}\ \freeify{a}} we obtain the value of the last write to address \isa{\freeify{a}} which is still pending in the store buffer. 
In case no outstanding write is in the store buffer we read from the memory. 
Store operations have no immediate effect on the memory but are queued in the store buffer instead. This also includes their effect on the ownership information.
Interlocked operations and the fence operation require an empty store buffer, which means that it has to be flushed before the action can take place. 
\begin{figure}
\begin{center}
\isa{\mbox{}\inferrule{\mbox{\freeify{v}\ {\isacharequal}\ {\isacharparenleft}\holkeyword{case}\ buffered{\isacharunderscore}val\ \freeify{sb}\ \freeify{a}\ \holkeyword{of}\ {\isasymbottom}\ {\isasymRightarrow}\ \freeify{m}\ \freeify{a}\ {\isacharbar}\ {\isasymlfloor}\boundify{v{\isacharprime}}{\isasymrfloor}\ {\isasymRightarrow}\ \boundify{v{\isacharprime}}{\isacharparenright}}\\\ \mbox{\freeify{sb{\isacharprime}}\ {\isacharequal}\ \freeify{sb}\ {\isacharat}\ {\isacharbrackleft}\constructor{Read\isactrlisub s\isactrlisub b}\ \freeify{volatile}\ \freeify{a}\ \freeify{t}\ \freeify{v}{\isacharbrackright}}}{\mbox{{\isacharparenleft}\constructor{Read}\ \freeify{volatile}\ \freeify{a}\ \freeify{t}\ {\isasymcdot}\ \freeify{is}{\isacharcomma}\ \freeify{{\isasymtheta}}{\isacharcomma}\ \freeify{sb}{\isacharcomma}\ \freeify{m}{\isacharcomma}\ \freeify{{\isasymG}}{\isacharparenright}\ $\overset{\isa{sb}}{\rightarrow}_{\isa{m}}$\ {\isacharparenleft}\freeify{is}{\isacharcomma}\ \freeify{{\isasymtheta}}{\isacharparenleft}\freeify{t}\ {\isasymmapsto}\ \freeify{v}{\isacharparenright}{\isacharcomma}\ \freeify{sb{\isacharprime}}{\isacharcomma}\ \freeify{m}{\isacharcomma}\ \freeify{{\isasymG}}{\isacharparenright}}}}\\[0.5\baselineskip]
\isa{\mbox{}\inferrule{\mbox{\freeify{sb{\isacharprime}}\ {\isacharequal}\ \freeify{sb}\ {\isacharat}\ {\isacharbrackleft}\constructor{Write\isactrlisub s\isactrlisub b}\ False\ \freeify{a}\ {\isacharparenleft}\freeify{D}{\isacharcomma}\ \freeify{f}{\isacharparenright}\ {\isacharparenleft}\freeify{f}\ \freeify{{\isasymtheta}}{\isacharparenright}\ \freeify{A}\ \freeify{L}\ \freeify{R}\ \freeify{W}{\isacharbrackright}}}{\mbox{{\isacharparenleft}\constructor{Write}\ False\ \freeify{a}\ {\isacharparenleft}\freeify{D}{\isacharcomma}\ \freeify{f}{\isacharparenright}\ \freeify{A}\ \freeify{L}\ \freeify{R}\ \freeify{W}\ {\isasymcdot}\ \freeify{is}{\isacharcomma}\ \freeify{{\isasymtheta}}{\isacharcomma}\ \freeify{sb}{\isacharcomma}\ \freeify{m}{\isacharcomma}\ \freeify{{\isasymG}}{\isacharparenright}\ $\overset{\isa{sb}}{\rightarrow}_{\isa{m}}$\ {\isacharparenleft}\freeify{is}{\isacharcomma}\ \freeify{{\isasymtheta}}{\isacharcomma}\ \freeify{sb{\isacharprime}}{\isacharcomma}\ \freeify{m}{\isacharcomma}\ \freeify{{\isasymG}}{\isacharparenright}}}}\\[0.5\baselineskip]
\isa{\mbox{}\inferrule{\mbox{\freeify{sb{\isacharprime}}\ {\isacharequal}\ \freeify{sb}\ {\isacharat}\ {\isacharbrackleft}\constructor{Write\isactrlisub s\isactrlisub b}\ True\ \freeify{a}\ {\isacharparenleft}\freeify{D}{\isacharcomma}\ \freeify{f}{\isacharparenright}\ {\isacharparenleft}\freeify{f}\ \freeify{{\isasymtheta}}{\isacharparenright}\ \freeify{A}\ \freeify{L}\ \freeify{R}\ \freeify{W}{\isacharbrackright}}\\\ \mbox{\freeify{{\isasymG}}\ {\isacharequal}\ {\isacharparenleft}\freeify{{\isasymD}}{\isacharcomma}\ \freeify{{\isasymO}}{\isacharcomma}\ \freeify{{\isasymA}}{\isacharcomma}\ \freeify{{\isasymS}}{\isacharparenright}}\\\ \mbox{\freeify{{\isasymG}{\isacharprime}}\ {\isacharequal}\ {\isacharparenleft}True{\isacharcomma}\ \freeify{{\isasymO}}{\isacharcomma}\ \freeify{{\isasymA}}{\isacharcomma}\ \freeify{{\isasymS}}{\isacharparenright}}}{\mbox{{\isacharparenleft}\constructor{Write}\ True\ \freeify{a}\ {\isacharparenleft}\freeify{D}{\isacharcomma}\ \freeify{f}{\isacharparenright}\ \freeify{A}\ \freeify{L}\ \freeify{R}\ \freeify{W}\ {\isasymcdot}\ \freeify{is}{\isacharcomma}\ \freeify{{\isasymtheta}}{\isacharcomma}\ \freeify{sb}{\isacharcomma}\ \freeify{m}{\isacharcomma}\ \freeify{{\isasymG}}{\isacharparenright}\ $\overset{\isa{sb}}{\rightarrow}_{\isa{m}}$\ {\isacharparenleft}\freeify{is}{\isacharcomma}\ \freeify{{\isasymtheta}}{\isacharcomma}\ \freeify{sb{\isacharprime}}{\isacharcomma}\ \freeify{m}{\isacharcomma}\ \freeify{{\isasymG}{\isacharprime}}{\isacharparenright}}}}\\[0.5\baselineskip]
\isa{\mbox{}\inferrule{\mbox{{\isasymnot}\ \freeify{cond}\ {\isacharparenleft}\freeify{{\isasymtheta}}{\isacharparenleft}\freeify{t}\ {\isasymmapsto}\ \freeify{m}\ \freeify{a}{\isacharparenright}{\isacharparenright}}\\\ \mbox{\freeify{{\isasymG}}\ {\isacharequal}\ {\isacharparenleft}\freeify{{\isasymD}}{\isacharcomma}\ \freeify{{\isasymO}}{\isacharcomma}\ \freeify{{\isasymA}}{\isacharcomma}\ \freeify{{\isasymS}}{\isacharparenright}}\\\ \mbox{\freeify{{\isasymG}{\isacharprime}}\ {\isacharequal}\ {\isacharparenleft}False{\isacharcomma}\ \freeify{{\isasymO}}{\isacharcomma}\ {\isasymemptyset}{\isacharcomma}\ \freeify{{\isasymS}}{\isacharparenright}}}{\mbox{{\isacharparenleft}\constructor{RMW}\ \freeify{a}\ \freeify{t}\ {\isacharparenleft}\freeify{D}{\isacharcomma}\ \freeify{f}{\isacharparenright}\ \freeify{cond}\ \freeify{ret}\ \freeify{A}\ \freeify{L}\ \freeify{R}\ \freeify{W}\ {\isasymcdot}\ \freeify{is}{\isacharcomma}\ \freeify{{\isasymtheta}}{\isacharcomma}\ {\isacharbrackleft}{\isacharbrackright}{\isacharcomma}\ \freeify{m}{\isacharcomma}\ \freeify{{\isasymG}}{\isacharparenright}\ $\overset{\isa{sb}}{\rightarrow}_{\isa{m}}$\ {\isacharparenleft}\freeify{is}{\isacharcomma}\ \freeify{{\isasymtheta}}{\isacharparenleft}\freeify{t}\ {\isasymmapsto}\ \freeify{m}\ \freeify{a}{\isacharparenright}{\isacharcomma}\ {\isacharbrackleft}{\isacharbrackright}{\isacharcomma}\ \freeify{m}{\isacharcomma}\ \freeify{{\isasymG}{\isacharprime}}{\isacharparenright}}}}\\[0.5\baselineskip]
\isa{\mbox{}\inferrule{\mbox{\freeify{cond}\ {\isacharparenleft}\freeify{{\isasymtheta}}{\isacharparenleft}\freeify{t}\ {\isasymmapsto}\ \freeify{m}\ \freeify{a}{\isacharparenright}{\isacharparenright}}\\\ \mbox{\freeify{{\isasymtheta}{\isacharprime}}\ {\isacharequal}\ \freeify{{\isasymtheta}}{\isacharparenleft}\freeify{t}\ {\isasymmapsto}\ \freeify{ret}\ {\isacharparenleft}\freeify{m}\ \freeify{a}{\isacharparenright}\ {\isacharparenleft}\freeify{f}\ {\isacharparenleft}\freeify{{\isasymtheta}}{\isacharparenleft}\freeify{t}\ {\isasymmapsto}\ \freeify{m}\ \freeify{a}{\isacharparenright}{\isacharparenright}{\isacharparenright}{\isacharparenright}}\\\ \mbox{\freeify{m{\isacharprime}}\ {\isacharequal}\ \freeify{m}{\isacharparenleft}\freeify{a}\ {\isacharcolon}{\isacharequal}\ \freeify{f}\ {\isacharparenleft}\freeify{{\isasymtheta}}{\isacharparenleft}\freeify{t}\ {\isasymmapsto}\ \freeify{m}\ \freeify{a}{\isacharparenright}{\isacharparenright}{\isacharparenright}}\\\ \mbox{\freeify{{\isasymG}}\ {\isacharequal}\ {\isacharparenleft}\freeify{{\isasymD}}{\isacharcomma}\ \freeify{{\isasymO}}{\isacharcomma}\ \freeify{{\isasymA}}{\isacharcomma}\ \freeify{{\isasymS}}{\isacharparenright}}\\\ \mbox{\freeify{{\isasymG}{\isacharprime}}\ {\isacharequal}\ {\isacharparenleft}False{\isacharcomma}\ \freeify{{\isasymO}}\ {\isasymunion}\ \freeify{A}\ {\isacharminus}\ \freeify{R}{\isacharcomma}\ {\isasymemptyset}{\isacharcomma}\ \freeify{{\isasymS}}\ {\isasymoplus}\isactrlbsub \freeify{W}\isactrlesub \ \freeify{R}\ {\isasymominus}\isactrlbsub \freeify{A}\isactrlesub \ \freeify{L}{\isacharparenright}}}{\mbox{{\isacharparenleft}\constructor{RMW}\ \freeify{a}\ \freeify{t}\ {\isacharparenleft}\freeify{D}{\isacharcomma}\ \freeify{f}{\isacharparenright}\ \freeify{cond}\ \freeify{ret}\ \freeify{A}\ \freeify{L}\ \freeify{R}\ \freeify{W}\ {\isasymcdot}\ \freeify{is}{\isacharcomma}\ \freeify{{\isasymtheta}}{\isacharcomma}\ {\isacharbrackleft}{\isacharbrackright}{\isacharcomma}\ \freeify{m}{\isacharcomma}\ \freeify{{\isasymG}}{\isacharparenright}\ $\overset{\isa{sb}}{\rightarrow}_{\isa{m}}$\ {\isacharparenleft}\freeify{is}{\isacharcomma}\ \freeify{{\isasymtheta}{\isacharprime}}{\isacharcomma}\ {\isacharbrackleft}{\isacharbrackright}{\isacharcomma}\ \freeify{m{\isacharprime}}{\isacharcomma}\ \freeify{{\isasymG}{\isacharprime}}{\isacharparenright}}}}\\[-0.3\baselineskip]
\isa{\mbox{}\inferrule{\mbox{}}{\mbox{{\isacharparenleft}\constructor{Fence}\ {\isasymcdot}\ \freeify{is}{\isacharcomma}\ \freeify{{\isasymtheta}}{\isacharcomma}\ {\isacharbrackleft}{\isacharbrackright}{\isacharcomma}\ \freeify{m}{\isacharcomma}\ \freeify{{\isasymD}}{\isacharcomma}\ \freeify{{\isasymO}}{\isacharcomma}\ \freeify{{\isasymA}}{\isacharcomma}\ \freeify{{\isasymS}}{\isacharparenright}\ $\overset{\isa{sb}}{\rightarrow}_{\isa{m}}$\ {\isacharparenleft}\freeify{is}{\isacharcomma}\ \freeify{{\isasymtheta}}{\isacharcomma}\ {\isacharbrackleft}{\isacharbrackright}{\isacharcomma}\ \freeify{m}{\isacharcomma}\ False{\isacharcomma}\ \freeify{{\isasymO}}{\isacharcomma}\ {\isasymemptyset}{\isacharcomma}\ \freeify{{\isasymS}}{\isacharparenright}}}}\\[-0.3\baselineskip]
\isa{\mbox{}\inferrule{\mbox{}}{\mbox{{\isacharparenleft}\constructor{Ghost}\ \freeify{A}\ \freeify{L}\ {\isasymcdot}\ \freeify{is}{\isacharcomma}\ \freeify{{\isasymtheta}}{\isacharcomma}\ \freeify{sb}{\isacharcomma}\ \freeify{m}{\isacharcomma}\ \freeify{{\isasymD}}{\isacharcomma}\ \freeify{{\isasymO}}{\isacharcomma}\ \freeify{{\isasymA}}{\isacharcomma}\ \freeify{{\isasymS}}{\isacharparenright}\ $\overset{\isa{sb}}{\rightarrow}_{\isa{m}}$\ {\isacharparenleft}\freeify{is}{\isacharcomma}\ \freeify{{\isasymtheta}}{\isacharcomma}\ \freeify{sb}\ {\isacharat}\ {\isacharbrackleft}\constructor{Ghost\isactrlisub s\isactrlisub b}\ \freeify{A}\ \freeify{L}{\isacharbrackright}{\isacharcomma}\ \freeify{m}{\isacharcomma}\ \freeify{{\isasymD}}{\isacharcomma}\ \freeify{{\isasymO}}{\isacharcomma}\ \freeify{{\isasymA}}{\isacharcomma}\ \freeify{{\isasymS}}{\isacharparenright}}}}
\end{center}
\caption{Memory transitions of store buffer machine\label{fig:store-buffer-memory}}
\end{figure}
We instantiate the global transition system with the rules of Figures \ref{fig:store-buffer-transitions} and \ref{fig:store-buffer-memory}. The program transitions are still a parameter. We refer to a transition by \isa{{\isacharparenleft}\freeify{ts}{\isacharcomma}\ \freeify{{\isasymS}}{\isacharcomma}\ \freeify{m}{\isacharparenright}\ $\overset{\isa{sb}}{\Rightarrow}$\ {\isacharparenleft}\freeify{ts{\isacharprime}}{\isacharcomma}\ \freeify{{\isasymS}{\isacharprime}}{\isacharcomma}\ \freeify{m{\isacharprime}}{\isacharparenright}}.%
\end{isamarkuptext}%
\isamarkuptrue%
\isamarkupsubsection{Coupling relation \label{sec:couplingrelation}%
}
\isamarkuptrue%
\begin{isamarkuptext}%
In this section we establish the coupling relation between a configuration of a machine with store buffer and the virtual machine without store buffer. 
It allows us to simulate every computation step of the store buffer machine by a sequence of steps (potentially empty) on the virtual machine. 
This transformation is essentially a sequentialization of the trace of the store buffer machine. 
When a thread of the store buffer machine executes a non-volatile operation, it only accesses memory which is not modified by any other thread (it is either owned or read-only). 
Although a non-volatile store is buffered, we can immediately execute it on the virtual machine, as there is no competing store of another thread.
The same is true for reads which get recorded in the store buffer.
However, with volatile writes we have to be careful, since concurrent threads may also compete with some volatile write to the same address. 
At the moment the volatile write enters the store buffer we do not yet know when it will be issued to memory and how it is ordered relatively to other outstanding writes of other threads.
We therefore suspend the write on the virtual machine from the moment it enters the store buffer to the moment it is issued to memory.
For volatile reads our access policy guarantees that there is no volatile write in the store buffer by flushing the store buffer if necessary. 
So there are at most some outstanding non-volatile writes in the store buffer, which are already executed on the virtual machine, as described before.
Altogether this suggests the following coupling relation: the memory of the virtual machine is obtained from the memory of the store buffer machine, by flushing every store buffer until we reach a volatile write. 
The remaining store buffer entries are suspended as instructions. 
The suspended reads are not yet visible in the temporaries of the virtual machine. 
Similar the ownership effects of the suspended ghost operations is not yet visible in the virtual machine.

Consider the following configuration of a thread \isa{\freeify{ts\isactrlisub s\isactrlisub b}\ensuremath{_{[\freeify{j}]}}} in the store buffer machine, where \isa{\freeify{i\isactrlisub k}} are the instructions and \isa{\freeify{s\isactrlisub k}} the store buffer entries. 
Let \isa{\freeify{s\isactrlisub v}} be the first volatile write in the store buffer. 
Keep in mind that new store buffer entries are appended to the end of the list and entries exit the store buffer and are issued to memory from the front of the list.
\begin{center} 
\isa{\freeify{ts\isactrlisub s\isactrlisub b}\ensuremath{_{[\freeify{j}]}}\ {\isacharequal}\ {\isacharparenleft}\freeify{p}{\isacharcomma}\ {\isacharbrackleft}\freeify{i\isactrlisub {\isadigit{1}}}{\isacharcomma}\ {\isasymdots}{\isacharcomma}\ \freeify{i\isactrlisub n}{\isacharbrackright}{\isacharcomma}\ \freeify{{\isasymtheta}}{\isacharcomma}\ {\isacharbrackleft}\freeify{s\isactrlisub {\isadigit{1}}}{\isacharcomma}\ {\isasymdots}{\isacharcomma}\ \freeify{s\isactrlisub v}{\isacharcomma}\ \freeify{s\isactrlisub {\isasymvv}}{\isacharcomma}\ {\isasymdots}{\isacharcomma}\ \freeify{s\isactrlisub m}{\isacharbrackright}{\isacharcomma}\ \freeify{{\isasymD}}{\isacharcomma}\ \freeify{{\isasymO}}{\isacharcomma}\ \freeify{{\isasymA}}{\isacharparenright}}
\end{center} 
The corresponding configuration \isa{\freeify{ts}\ensuremath{_{[\freeify{j}]}}} in the virtual machine is obtained by suspending all store buffer entries beginning at \isa{\freeify{s\isactrlisub v}} to the front of the instructions. 
A store buffer \isa{\constructor{Read\isactrlisub s\isactrlisub b}} / \isa{\constructor{Write\isactrlisub s\isactrlisub b}} / \isa{\constructor{Ghost\isactrlisub s\isactrlisub b}} is converted to a  \isa{\constructor{Read}} / \isa{\constructor{Write}} / \isa{\constructor{Ghost}} instruction. 
We take the freedom to make this coercion implicit in the example. 
The store buffer entries preceding \isa{\freeify{s\isactrlisub v}} have already made their way to memory, whereas the suspended read operations are not yet visible in the temporaries \isa{\freeify{{\isasymtheta}{\isacharprime}}}. Similar, the suspended updates to the ownership sets and dirty flag are not yet recorded in \isa{\freeify{{\isasymO}{\isacharprime}}}, \isa{\freeify{{\isasymA}{\isacharprime}}} and \isa{\freeify{{\isasymD}{\isacharprime}}}.
\begin{center} 
\isa{\freeify{ts}\ensuremath{_{[\freeify{j}]}}\ {\isacharequal}\ {\isacharparenleft}\freeify{p}{\isacharcomma}\ {\isacharbrackleft}\freeify{s\isactrlisub v}{\isacharcomma}\ \freeify{s\isactrlisub {\isasymvv}}{\isacharcomma}\ {\isasymdots}{\isacharcomma}\ \freeify{s\isactrlisub m}{\isacharcomma}\ \freeify{i\isactrlisub {\isadigit{1}}}{\isacharcomma}\ {\isasymdots}{\isacharcomma}\ \freeify{i\isactrlisub n}{\isacharbrackright}{\isacharcomma}\ \freeify{{\isasymtheta}{\isacharprime}}{\isacharcomma}\ {\isacharparenleft}{\isacharparenright}{\isacharcomma}\ \freeify{{\isasymD}{\isacharprime}}{\isacharcomma}\ \freeify{{\isasymO}{\isacharprime}}{\isacharcomma}\ \freeify{{\isasymA}{\isacharprime}}{\isacharparenright}}
\end{center} 
This example illustrates that the virtual machine falls behind the store buffer machine in our simulation, as store buffer instructions are suspended and reads (and ghost operations) are delayed and not yet visible in the temporaries (and the ghost state).
This delay can also propagate to the level of the programming language, which communicates with the memory system by reading the temporaries and issuing new instructions. 
For example the control flow can depend on the temporaries, which store the result of branching conditions. 
It may happen that the store buffer machine already has evaluated the branching condition by referring to the values in the store buffer, whereas the virtual machine still has to wait. 
Formally this manifests in still undefined temporaries. 
Now consider that the program in the store buffer machine makes a step from \isa{\freeify{p}} to \isa{{\isacharparenleft}\freeify{p{\isacharprime}}{\isacharcomma}\ \freeify{is{\isacharprime}}{\isacharparenright}}, which results in a thread configuration where the program state has switched to \isa{\freeify{p{\isacharprime}}}, the instructions \isa{\freeify{is{\isacharprime}}} are appended and the program step is recorded in the store buffer:
\begin{center} 
\isa{\freeify{ts\isactrlisub s\isactrlisub b{\isacharprime}}\ensuremath{_{[\freeify{j}]}}\ {\isacharequal}\ {\isacharparenleft}\freeify{p{\isacharprime}}{\isacharcomma}\ {\isacharbrackleft}\freeify{i\isactrlisub {\isadigit{1}}}{\isacharcomma}\ {\isasymdots}{\isacharcomma}\ \freeify{i\isactrlisub n}{\isacharbrackright}\ {\isacharat}\ \freeify{is{\isacharprime}}{\isacharcomma}\ \freeify{{\isasymtheta}}{\isacharcomma}\ {\isacharbrackleft}\freeify{s\isactrlisub {\isadigit{1}}}{\isacharcomma}\ {\isasymdots}{\isacharcomma}\ \freeify{s\isactrlisub v}{\isacharcomma}\ {\isasymdots}{\isacharcomma}\ \freeify{s\isactrlisub m}{\isacharcomma}\ \constructor{Prog\isactrlisub s\isactrlisub b}\ \freeify{p}\ \freeify{p{\isacharprime}}\ \freeify{is{\isacharprime}}{\isacharbrackright}{\isacharcomma}\ \freeify{{\isasymD}}{\isacharcomma}\ \freeify{{\isasymO}}{\isacharcomma}\ \freeify{{\isasymA}}{\isacharparenright}}
\end{center} 
The virtual machine however makes no step, since it still has to evaluate the suspended instructions before making the program step. 
The instructions \isa{\freeify{is{\isacharprime}}} are not yet issued and the program state is still \isa{\freeify{p}}. 
We also take these program steps into account in our final coupling relation \isa{{\isacharparenleft}\freeify{ts\isactrlisub s\isactrlisub b}{\isacharcomma}\ \freeify{{\isasymS}\isactrlisub s\isactrlisub b}{\isacharcomma}\ \freeify{m\isactrlisub s\isactrlisub b}{\isacharparenright}\ {\isasymsim}\ {\isacharparenleft}\freeify{ts}{\isacharcomma}\ \freeify{{\isasymS}}{\isacharcomma}\ \freeify{m}{\isacharparenright}}, defined in Figure~\ref{fig:coupling-relation}.
\begin{figure}
\begin{center}
\begin{minipage}{10cm}
\inferrule{\isa{\freeify{m}\ {\isacharequal}\ flush{\isacharunderscore}all{\isacharunderscore}until{\isacharunderscore}volatile{\isacharunderscore}write\ \freeify{ts\isactrlisub s\isactrlisub b}\ \freeify{m\isactrlisub s\isactrlisub b}}\\
           \isa{\freeify{{\isasymS}}\ {\isacharequal}\ share{\isacharunderscore}all{\isacharunderscore}until{\isacharunderscore}volatile{\isacharunderscore}write\ \freeify{ts\isactrlisub s\isactrlisub b}\ \freeify{{\isasymS}\isactrlisub s\isactrlisub b}}\\
           \isa{{\isacharbar}\freeify{ts\isactrlisub s\isactrlisub b}{\isacharbar}\ {\isacharequal}\ {\isacharbar}\freeify{ts}{\isacharbar}}\\
           \parbox{9.8cm}{\isa{{\isasymforall}\boundify{i}{\isacharless}{\isacharbar}\freeify{ts\isactrlisub s\isactrlisub b}{\isacharbar}{\isachardot}\isanewline
\isaindent{\ \ \ }\holkeyword{let}\ {\isacharparenleft}\boundify{p}{\isacharcomma}\ \boundify{is\isactrlisub s\isactrlisub b}{\isacharcomma}\ \boundify{{\isasymtheta}}{\isacharcomma}\ \boundify{sb}{\isacharcomma}\ \boundify{{\isasymD}\isactrlisub s\isactrlisub b}{\isacharcomma}\ \boundify{{\isasymO}}{\isacharcomma}\ \boundify{{\isasymA}}{\isacharparenright}\ {\isacharequal}\ \freeify{ts\isactrlisub s\isactrlisub b}\ensuremath{_{[\boundify{i}]}}{\isacharsemicolon}\isanewline
\isaindent{\ \ \ \holkeyword{let}\ }\boundify{flushs}\ {\isacharequal}\ takeWhile\ not{\isacharunderscore}volatile{\isacharunderscore}write\ \boundify{sb}{\isacharsemicolon}\isanewline
\isaindent{\ \ \ \holkeyword{let}\ }\boundify{suspends}\ {\isacharequal}\ dropWhile\ not{\isacharunderscore}volatile{\isacharunderscore}write\ \boundify{sb}\isanewline
\isaindent{\ \ \ }\holkeyword{in}\ {\isasymexists}\boundify{is}\ \boundify{{\isasymD}}{\isachardot}\ instrs\ \boundify{suspends}\ {\isacharat}\ \boundify{is\isactrlisub s\isactrlisub b}\ {\isacharequal}\ \boundify{is}\ {\isacharat}\ prog{\isacharunderscore}instrs\ \boundify{suspends}\ {\isasymand}\isanewline
\isaindent{\ \ \ \holkeyword{in}\ {\isasymexists}\boundify{is}\ \boundify{{\isasymD}}{\isachardot}\ }\boundify{{\isasymD}\isactrlisub s\isactrlisub b}\ {\isacharequal}\ {\isacharparenleft}\boundify{{\isasymD}}\ {\isasymor}\ refs\ volatile{\isacharunderscore}Write\ \boundify{sb}\ {\isasymnoteq}\ {\isasymemptyset}{\isacharparenright}\ {\isasymand}\isanewline
\isaindent{\ \ \ \holkeyword{in}\ {\isasymexists}\boundify{is}\ \boundify{{\isasymD}}{\isachardot}\ }\freeify{ts}\ensuremath{_{[\boundify{i}]}}\ {\isacharequal}\isanewline
\isaindent{\ \ \ \holkeyword{in}\ {\isasymexists}\boundify{is}\ \boundify{{\isasymD}}{\isachardot}\ }{\isacharparenleft}hd{\isacharunderscore}prog\ \boundify{p}\ \boundify{suspends}{\isacharcomma}\ \boundify{is}{\isacharcomma}\ \boundify{{\isasymtheta}}{\isasymrestriction}\isactrlbsub {\isacharparenleft}{\isacharminus}\ read{\isacharunderscore}tmps\ \boundify{suspends}{\isacharparenright}\isactrlesub {\isacharcomma}\ {\isacharparenleft}{\isacharparenright}{\isacharcomma}\ \boundify{{\isasymD}}{\isacharcomma}\isanewline
\isaindent{\ \ \ \holkeyword{in}\ {\isasymexists}\boundify{is}\ \boundify{{\isasymD}}{\isachardot}\ \ }acquire\ \boundify{flushs}\ \boundify{{\isasymO}}{\isacharcomma}\ acquire\ \boundify{flushs}\ \boundify{{\isasymA}}{\isacharparenright}}}}
          {\isa{{\isacharparenleft}\freeify{ts\isactrlisub s\isactrlisub b}{\isacharcomma}\ \freeify{{\isasymS}\isactrlisub s\isactrlisub b}{\isacharcomma}\ \freeify{m\isactrlisub s\isactrlisub b}{\isacharparenright}\ {\isasymsim}\ {\isacharparenleft}\freeify{ts}{\isacharcomma}\ \freeify{{\isasymS}}{\isacharcomma}\ \freeify{m}{\isacharparenright}}}
\end{minipage}
\end{center}
\caption{Coupling relation \label{fig:coupling-relation}}
\end{figure}
We denote the already simulated (flushed) store buffer entries by \isa{\boundify{flushs}} and the suspended ones by \isa{\boundify{suspends}}. 
The function \isa{instrs} converts them back to instructions, which are a prefix of the instructions of the virtual machine. 
We collect the additional instructions which were issued by program instructions but still recorded in the remainder of the store buffer with function \isa{prog{\isacharunderscore}instrs}. 
These instructions have already made their way to the instructions of the store buffer machine but not yet on the virtual machine. 
This situation is formalized as \isa{instrs\ \boundify{suspends}\ {\isacharat}\ \boundify{is\isactrlisub s\isactrlisub b}\ {\isacharequal}\ \boundify{is}\ {\isacharat}\ prog{\isacharunderscore}instrs\ \boundify{suspends}}, where \isa{\boundify{is}} are the instructions of the virtual machine. 
The program state of the virtual machine is either the same as in the store buffer machine or the first program state recorded in the suspended part of the store buffer.
This state is selected by \isa{hd{\isacharunderscore}prog}. 
The temporaries of the virtual machine are obtained by removing the suspended reads from \isa{\freeify{{\isasymtheta}}}. 
The memory is obtained by flushing all store buffers until the first volatile write is hit, excluding it. Thereby only non-volatile writes are flushed, which are all thread local, and hence could be flushed in any order with the same result on the memory. Function \isa{flush{\isacharunderscore}all{\isacharunderscore}until{\isacharunderscore}volatile{\isacharunderscore}write} flushes them in order of appearance.
Similarly the sharing map of the virtual machine is obtained by flushing all store buffers until the first volatile write via the function \isa{share{\isacharunderscore}all{\isacharunderscore}until{\isacharunderscore}volatile{\isacharunderscore}write}. For the local ownership sets \isa{\freeify{{\isasymO}}} and \isa{\freeify{{\isasymA}}} the auxiliary function \isa{acquire} calculates the outstanding effect of the already simulated parts of the store buffer.

One may think of simplifying the coupling relation by avoiding flushing altogether and just suspending the whole store buffer. However, consider the following scenario. 
A thread is reading from a volatile address. 
It can still have non-volatile writes in its store buffer. 
Hence the read would be suspended, and we could miss updates made by other threads to this address. 
\end{isamarkuptext}%
\isamarkuptrue%
\isamarkupsubsection{Simulation \label{sec:simulation}%
}
\isamarkuptrue%
\begin{isamarkuptext}%
Theorem \ref{thm:simulation} is our core simulation theorem. 
Provided that all reachable states of the virtual machine are safe, a step of the store buffer machine can be simulated by a (potentially empty) sequence of steps on the virtual machine, maintaining the coupling relation and an invariant on the configurations of the store buffer machine.
\begin{theorem}[Simulation]\label{thm:simulation}
\begin{isabelle}%
{\isacharparenleft}\freeify{ts\isactrlisub s\isactrlisub b}{\isacharcomma}\ \freeify{{\isasymS}\isactrlisub s\isactrlisub b}{\isacharcomma}\ \freeify{m\isactrlisub s\isactrlisub b}{\isacharparenright}\ $\overset{\isa{sb}}{\Rightarrow}$\ {\isacharparenleft}\freeify{ts\isactrlisub s\isactrlisub b{\isacharprime}}{\isacharcomma}\ \freeify{{\isasymS}\isactrlisub s\isactrlisub b{\isacharprime}}{\isacharcomma}\ \freeify{m\isactrlisub s\isactrlisub b{\isacharprime}}{\isacharparenright}\ {\isasymand}\ {\isacharparenleft}\freeify{ts\isactrlisub s\isactrlisub b}{\isacharcomma}\ \freeify{{\isasymS}\isactrlisub s\isactrlisub b}{\isacharcomma}\ \freeify{m\isactrlisub s\isactrlisub b}{\isacharparenright}\ {\isasymsim}\ {\isacharparenleft}\freeify{ts}{\isacharcomma}\ \freeify{{\isasymS}}{\isacharcomma}\ \freeify{m}{\isacharparenright}\ {\isasymand}\isanewline
\isaindent{\ \ }safe{\isacharunderscore}reach\ {\isacharparenleft}\freeify{ts}{\isacharcomma}\ \freeify{{\isasymS}}{\isacharcomma}\ \freeify{m}{\isacharparenright}\ {\isasymand}\ invariant\ \freeify{ts\isactrlisub s\isactrlisub b}\ \freeify{{\isasymS}\isactrlisub s\isactrlisub b}\ \freeify{m\isactrlisub s\isactrlisub b}\ {\isasymlongrightarrow}\isanewline
\isaindent{\ \ }\ invariant\ \freeify{ts\isactrlisub s\isactrlisub b{\isacharprime}}\ \freeify{{\isasymS}\isactrlisub s\isactrlisub b{\isacharprime}}\ \freeify{m\isactrlisub s\isactrlisub b{\isacharprime}}\ {\isasymand}\isanewline
\isaindent{\ \ }{\isacharparenleft}{\isasymexists}\boundify{ts{\isacharprime}}\ \boundify{{\isasymS}{\isacharprime}}\ \boundify{m{\isacharprime}}{\isachardot}\ {\isacharparenleft}\freeify{ts}{\isacharcomma}\ \freeify{{\isasymS}}{\isacharcomma}\ \freeify{m}{\isacharparenright}\ $\overset{\isa{v}}{\Rightarrow}^{*}$\ {\isacharparenleft}\boundify{ts{\isacharprime}}{\isacharcomma}\ \boundify{{\isasymS}{\isacharprime}}{\isacharcomma}\ \boundify{m{\isacharprime}}{\isacharparenright}\ {\isasymand}\ {\isacharparenleft}\freeify{ts\isactrlisub s\isactrlisub b{\isacharprime}}{\isacharcomma}\ \freeify{{\isasymS}\isactrlisub s\isactrlisub b{\isacharprime}}{\isacharcomma}\ \freeify{m\isactrlisub s\isactrlisub b{\isacharprime}}{\isacharparenright}\ {\isasymsim}\ {\isacharparenleft}\boundify{ts{\isacharprime}}{\isacharcomma}\ \boundify{{\isasymS}{\isacharprime}}{\isacharcomma}\ \boundify{m{\isacharprime}}{\isacharparenright}{\isacharparenright}%
\end{isabelle}
\end{theorem}
In the following we discuss the invariant \isa{invariant\ \freeify{ts\isactrlisub s\isactrlisub b}\ \freeify{S}\ \freeify{m\isactrlisub s\isactrlisub b}}, where we commonly refer to a thread configuration \isa{\freeify{ts\isactrlisub s\isactrlisub b}\ensuremath{_{[\freeify{i}]}}\ {\isacharequal}\ {\isacharparenleft}\freeify{p}{\isacharcomma}\ \freeify{is}{\isacharcomma}\ \freeify{{\isasymtheta}}{\isacharcomma}\ \freeify{sb}{\isacharcomma}\ \freeify{{\isasymD}}{\isacharcomma}\ \freeify{{\isasymO}}{\isacharcomma}\ \freeify{{\isasymA}}{\isacharparenright}} for \isa{\freeify{i}\ {\isacharless}\ {\isacharbar}\freeify{ts\isactrlisub s\isactrlisub b}{\isacharbar}}. 
By outstanding references we refer to read and write operations in the store buffer. 
The invariant is a conjunction of several sub-invariants grouped by their content:

\begin{isabelle}%
invariant\ \freeify{ts\isactrlisub s\isactrlisub b}\ \freeify{S}\ \freeify{m\isactrlisub s\isactrlisub b}\ {\isasymequiv}\ ownership{\isacharunderscore}inv\ \freeify{S}\ \freeify{ts\isactrlisub s\isactrlisub b}\ {\isasymand}\ sharing{\isacharunderscore}inv\ \freeify{S}\ \freeify{ts\isactrlisub s\isactrlisub b}\ {\isasymand}\isanewline
\isaindent{\ \ }temporaries{\isacharunderscore}inv\ \freeify{ts\isactrlisub s\isactrlisub b}\ {\isasymand}\ data{\isacharunderscore}dependency{\isacharunderscore}inv\ \freeify{ts\isactrlisub s\isactrlisub b}\ {\isasymand}\ history{\isacharunderscore}inv\ \freeify{ts\isactrlisub s\isactrlisub b}\ \freeify{m\isactrlisub s\isactrlisub b}\ {\isasymand}\isanewline
\isaindent{\ \ }flush{\isacharunderscore}inv\ \freeify{ts\isactrlisub s\isactrlisub b}\ {\isasymand}\ \freeify{valid}\ \freeify{ts\isactrlisub s\isactrlisub b}%
\end{isabelle}

\paragraph{Ownership.} 
\begin{inparaenum}
\item For every thread all outstanding non-volatile references have to be owned or refer to read-only memory.
\item Every outstanding volatile write is not owned by any other thread. 
\item Outstanding accesses to read-only memory are not owned.
\item The ownership sets of every two different threads are distinct.
\end{inparaenum}

\paragraph{Sharing.}
\begin{inparaenum}
\item All outstanding non volatile writes are unshared. 
\item All unowned addresses are shared.
\item No thread owns read-only memory.
\item The ownership annotations of outstanding ghost and write operations are consistent (\eg released addresses are owned at the point of release).
\item There is no outstanding write to read-only memory.
\end{inparaenum}

\paragraph{Temporaries.} Temporaries are modeled as an unlimited store for temporary registers. We require certain distinctness and freshness properties for each thread.
\begin{inparaenum}
\item The temporaries referred to by read instructions are distinct.
\item The temporaries referred to by reads in the store buffer are distinct.
\item Read and write temporaries are distinct.
\item Read temporaries are fresh, \ie are not in the domain of \isa{\freeify{{\isasymtheta}}}.
\end{inparaenum}

\paragraph{Data dependency.} Data dependency means that store operations may only depend on \emph{previous} read operations. For every thread we have:
\begin{inparaenum}
\item Every operation \isa{{\isacharparenleft}\freeify{D}{\isacharcomma}\ \freeify{f}{\isacharparenright}} in a write instruction or a store buffer write is valid according to \isa{valid{\isacharunderscore}sop\ {\isacharparenleft}\freeify{D}{\isacharcomma}\ \freeify{f}{\isacharparenright}}, \ie function \isa{\freeify{f}} only depends on domain \isa{\freeify{D}}.
\item For every suffix of the instructions of the form \isa{\constructor{Write}\ \freeify{volatile}\ \freeify{a}\ {\isacharparenleft}\freeify{D}{\isacharcomma}\ \freeify{f}{\isacharparenright}\ \freeify{A}\ \freeify{L}\ \freeify{R}\ \freeify{W}\ {\isasymcdot}\ \freeify{is}} the domain \isa{\freeify{D}} is distinct from the temporaries referred to by future read instructions in \isa{\freeify{is}}.
\item The outstanding writes in the store buffer do not depend on the read temporaries still in the instruction list.
\end{inparaenum}

\paragraph{History.} The history information  of program steps and  read operations we record in the store buffer have to be consistent with the trace. For every thread:
\begin{inparaenum}
\item The value stored for a non volatile read is the same as the last write to the same address in the store buffer or the value in memory, in case there is no write in the buffer. 
\item All reads have to be clean. This results from our flushing policy. Note that the value recorded for a volatile (and acquired non-volatile) read in the initial part of the store buffer (before the first volatile write), may become stale with respect to the memory. Remember that those parts of the store buffer are already flushed in the virtual machine and thus cause no trouble.
\item For every read the recorded value coincides with the corresponding value in the temporaries.
\item For every \isa{\constructor{Write\isactrlisub s\isactrlisub b}\ \freeify{volatile}\ \freeify{a}\ {\isacharparenleft}\freeify{D}{\isacharcomma}\ \freeify{f}{\isacharparenright}\ \freeify{v}\ \freeify{A}\ \freeify{L}\ \freeify{R}\ \freeify{W}} the recorded value \isa{\freeify{v}} coincides with \isa{\freeify{f}\ \freeify{{\isasymtheta}}}, and domain \isa{\freeify{D}} is subset of \isa{dom\ \freeify{{\isasymtheta}}} and is distinct from the following read temporaries. Note that the consistency of the ownership annotations is already covered by the aforementioned invariants.
\item For every suffix in the store buffer of the form \isa{\constructor{Prog\isactrlisub s\isactrlisub b}\ \freeify{p\isactrlisub {\isadigit{1}}}\ \freeify{p\isactrlisub {\isadigit{2}}}\ \freeify{is{\isacharprime}}\ {\isasymcdot}\ \freeify{sb{\isacharprime}}}, either \isa{\freeify{p\isactrlisub {\isadigit{1}}}\ {\isacharequal}\ \freeify{p}} in case there is no preceding program node in the buffer or it corresponds to the last program state recorded there. 
Moreover, the program transition \isa{\freeify{{\isasymtheta}}{\isasymrestriction}\isactrlbsub {\isacharparenleft}{\isacharminus}\ read{\isacharunderscore}tmps\ \freeify{sb{\isacharprime}}{\isacharparenright}\isactrlesub {\isasymturnstile}\ \freeify{p\isactrlisub {\isadigit{1}}}\ {\isasymrightarrow}\isactrlsub p\ {\isacharparenleft}\freeify{p\isactrlisub {\isadigit{2}}}{\isacharcomma}\ \freeify{is{\isacharprime}}{\isacharparenright}} is possible, \ie it was possible to execute the program transition at that point.
\item The program configuration \isa{\freeify{p}} coincides with the last program configuration recorded in the store buffer.
\item As the instructions from a program step are at the one hand appended to the instruction list and on the other hand recorded in the store buffer, we have for every suffix \isa{\freeify{sb{\isacharprime}}} of the store buffer: \isa{{\isasymexists}\boundify{is{\isacharprime}}{\isachardot}\ instrs\ \freeify{sb{\isacharprime}}\ {\isacharat}\ \freeify{is}\ {\isacharequal}\ \boundify{is{\isacharprime}}\ {\isacharat}\ prog{\isacharunderscore}instrs\ \freeify{sb{\isacharprime}}}, \ie the remaining instructions \isa{\freeify{is}} correspond to a suffix of the recorded instructions \isa{prog{\isacharunderscore}instrs\ \freeify{sb{\isacharprime}}}.
\end{inparaenum}

\paragraph{Flushes.} If the dirty flag is unset there are no outstanding volatile writes in the store buffer.

\paragraph{Program step.} The program-transitions are still a parameter of our model. 
In order to make the proof work, we have to assume some of the invariants also for the program steps. 
We allow the program-transitions to employ further invariants on the configurations, these are modeled by the parameter \isa{\freeify{valid}}. 
For example, in the instantiation later on the program keeps a counter for the temporaries, for each thread. 
We maintain distinctness of temporaries by restricting all temporaries occurring in the memory system to be below that counter, which is expressed by instantiating \isa{\freeify{valid}}. 
Program steps, memory steps and store buffer steps have to maintain \isa{\freeify{valid}}. 
Furthermore we assume the following properties of a program step:
\begin{inparaenum}
\item The program step generates fresh, distinct read temporaries, that are neither in \isa{\freeify{{\isasymtheta}}} nor in the store buffer temporaries of the memory system.
\item The generated memory instructions respect data dependencies, and are valid according to \isa{valid{\isacharunderscore}sop}.

\end{inparaenum}

\paragraph{Proof.} We do not go into details but rather sketch the main arguments for simulation of a step in the store buffer machine by a potentially empty sequence of steps in the virtual machine, maintaining the coupling relation. 
The first case distinction in the proof is on the global transitions in Figure~\ref{fig:global-transitions}. 
\begin{inparaenum}
\item \emph{Program step}: 
we make a case distinction whether there is an outstanding volatile write in the store buffer or not. 
If not the configuration of the virtual machine corresponds to the flushed store buffer and we can make the same step. 
Otherwise the virtual machine makes no step as we have to wait until all volatile writes have exited the store buffer.
\item \emph{Memory step}: 
we do case distinction on the rules in Figure~\ref{fig:store-buffer-memory}. 
For read, non volatile write and ghost instructions we do the same case distinction as for the program step. 
If there is no outstanding volatile write in the store buffer we can make the step, otherwise we have to wait. 
When a volatile write enters the store buffer it is suspended until it exists the store buffer. Hence we do no step in the virtual machine.
The read-modify-write and the fence instruction can all be simulated immediately since the store buffer has to be empty.
\item \emph{Store Buffer step}:
we do case distinction on the rules in Figure~\ref{fig:store-buffer-transitions}. 
When a read, a non volatile write, a ghost operation or a program history node exits the store buffer, the virtual machine does not have to do any step since these steps are already visible. 
When a volatile write exits the store buffer, we execute all the suspended operations (including reads, ghost operations and program steps) until the next suspended volatile write is hit. This is possible since all writes are non volatile and thus memory modifications are thread local. 
\end{inparaenum}

A common argument in various places in the proof is to rule out potential races by constructing calculations of the virtual machine that lead to an unsafe state and are thus unreachable in a safe execution.
Here we make use of the fact, that the ghost operations in the prefixes of the store buffers that are already simulated in the virtual machine may only acquire new addresses to the ownership of a thread but not realease addresses. 
From the viewpoint of other threads this may only lead to a more restrictive configuration but never to more liberal one, making the construction of an unsafe execution of the virtual machine possible without referring to an older state as the current state of the store buffer machine.%
\end{isamarkuptext}%
\isamarkuptrue%
\isamarkupsubsection{PIMP \label{sec:pimp}%
}
\isamarkuptrue%
\begin{isamarkuptext}%
PIMP is a parallel version of IMP\cite{Nipkow-FSTTCS-96}, a canonical WHILE-language.

An expression \isa{\freeify{e}} is either 
\begin{inparaenum}
\item \isa{\constructor{Const}\ \freeify{v}}, a constant value,
\item \isa{\constructor{Mem}\ \freeify{volatile}\ \freeify{a}}, a (volatile) memory lookup at address \isa{\freeify{a}},
\item \isa{\constructor{Tmp}\ \freeify{sop}}, reading from the temporaries with a operation \isa{\freeify{sop}} which is an intermediate expression occurring in the transition rules for statements,
\item \isa{\constructor{Unop}\ \freeify{f}\ \freeify{e}}, a unary operation where \isa{\freeify{f}} is a unary function on values, and finally
\item \isa{\constructor{Binop}\ \freeify{f}\ \freeify{e\isactrlisub {\isadigit{1}}}\ \freeify{e\isactrlisub {\isadigit{2}}}}, a binary operation where \isa{\freeify{f}} is a binary function on values.
\end{inparaenum}

A statement \isa{\freeify{s}} is either
\begin{inparaenum}
\item \isa{\constructor{Skip}}, the empty statement,
\item \isa{\constructor{Assign}\ \freeify{volatile}\ \freeify{a}\ \freeify{e}\ \freeify{A}\ \freeify{L}\ \freeify{R}\ \freeify{W}}, a (volatile) assignment of expression \isa{\freeify{e}} to address expression \isa{\freeify{a}},
\item \isa{\constructor{CAS}\ \freeify{a}\ \freeify{c\isactrlisub e}\ \freeify{s\isactrlisub e}\ \freeify{A}\ \freeify{L}\ \freeify{R}\ \freeify{W}}, atomic compare and swap at address expression \isa{\freeify{a}} with compare expression 
  \isa{\freeify{c\isactrlisub e}} and swap expression \isa{\freeify{s\isactrlisub e}},
\item \isa{\constructor{Seq}\ \freeify{s\isactrlisub {\isadigit{1}}}\ \freeify{s\isactrlisub {\isadigit{2}}}}, sequential composition,
\item \isa{\constructor{Cond}\ \freeify{e}\ \freeify{s\isactrlisub {\isadigit{1}}}\ \freeify{s\isactrlisub {\isadigit{2}}}}, the if-then-else statement,
\item \isa{\constructor{While}\ \freeify{e}\ \freeify{s}}, the loop statement with condition \isa{\freeify{e}},
\item \isa{\constructor{SGhost}}, and \isa{\constructor{SFence}} as stubs for the corresponding memory instructions.
\end{inparaenum}

The key idea of the semantics is the following: expressions are evaluated by issuing instructions to the memory system, then the program waits until the memory system has made all necessary results available in the temporaries, which allows the program to make another step. Figure~\ref{fig:expression-evaluation} defines expression evaluation.
\begin{figure}
\begin{tabularx}{\textwidth}{l@ {~~\isa{{\isacharequal}}~~}X}
\isa{issue{\isacharunderscore}expr\ \freeify{t}\ {\isacharparenleft}\constructor{Const}\ \freeify{v}{\isacharparenright}} & \isa{{\isacharbrackleft}{\isacharbrackright}}\\
\isa{issue{\isacharunderscore}expr\ \freeify{t}\ {\isacharparenleft}\constructor{Mem}\ \freeify{volatile}\ \freeify{a}{\isacharparenright}} & \isa{{\isacharbrackleft}\constructor{Read}\ \freeify{volatile}\ \freeify{a}\ \freeify{t}{\isacharbrackright}}\\
\isa{issue{\isacharunderscore}expr\ \freeify{t}\ {\isacharparenleft}\constructor{Tmp}\ {\isacharparenleft}\freeify{D}{\isacharcomma}\ \freeify{f}{\isacharparenright}{\isacharparenright}} & \isa{{\isacharbrackleft}{\isacharbrackright}}\\
\isa{issue{\isacharunderscore}expr\ \freeify{t}\ {\isacharparenleft}\constructor{Unop}\ \freeify{f}\ \freeify{e}{\isacharparenright}} & \isa{issue{\isacharunderscore}expr\ \freeify{t}\ \freeify{e}}\\
\isa{issue{\isacharunderscore}expr\ \freeify{t}\ {\isacharparenleft}\constructor{Binop}\ \freeify{f}\ \freeify{e\isactrlisub {\isadigit{1}}}\ \freeify{e\isactrlisub {\isadigit{2}}}{\isacharparenright}} & \isa{issue{\isacharunderscore}expr\ \freeify{t}\ \freeify{e\isactrlisub {\isadigit{1}}}\ {\isacharat}\ issue{\isacharunderscore}expr\ {\isacharparenleft}\freeify{t}\ {\isacharplus}\ used{\isacharunderscore}tmps\ \freeify{e\isactrlisub {\isadigit{1}}}{\isacharparenright}\ \freeify{e\isactrlisub {\isadigit{2}}}}\\
\end{tabularx}\\[2pt]

\begin{tabularx}{\textwidth}{l@ {~~\isa{{\isacharequal}}~~}X}
\isa{eval{\isacharunderscore}expr\ \freeify{t}\ {\isacharparenleft}\constructor{Const}\ \freeify{v}{\isacharparenright}} & \isa{{\isacharparenleft}{\isasymemptyset}{\isacharcomma}\ {\isasymlambda}\boundify{{\isasymtheta}}{\isachardot}\ \freeify{v}{\isacharparenright}}\\
\isa{eval{\isacharunderscore}expr\ \freeify{t}\ {\isacharparenleft}\constructor{Mem}\ \freeify{volatile}\ \freeify{a}{\isacharparenright}} & \isa{{\isacharparenleft}{\isacharbraceleft}\freeify{t}{\isacharbraceright}{\isacharcomma}\ {\isasymlambda}\boundify{{\isasymtheta}}{\isachardot}\ the\ {\isacharparenleft}\boundify{{\isasymtheta}}\ \freeify{t}{\isacharparenright}{\isacharparenright}}\\
\isa{eval{\isacharunderscore}expr\ \freeify{t}\ {\isacharparenleft}\constructor{Tmp}\ {\isacharparenleft}\freeify{D}{\isacharcomma}\ \freeify{f}{\isacharparenright}{\isacharparenright}} & \isa{{\isacharparenleft}\freeify{D}{\isacharcomma}\ \freeify{f}{\isacharparenright}}\\
\isa{eval{\isacharunderscore}expr\ \freeify{t}\ {\isacharparenleft}\constructor{Unop}\ \freeify{f}\ \freeify{e}{\isacharparenright}} & \isa{\holkeyword{let}\ {\isacharparenleft}\boundify{D}{\isacharcomma}\ \boundify{f\isactrlisub e}{\isacharparenright}\ {\isacharequal}\ eval{\isacharunderscore}expr\ \freeify{t}\ \freeify{e}\ \holkeyword{in}\ {\isacharparenleft}\boundify{D}{\isacharcomma}\ {\isasymlambda}\boundify{{\isasymtheta}}{\isachardot}\ \freeify{f}\ {\isacharparenleft}\boundify{f\isactrlisub e}\ \boundify{{\isasymtheta}}{\isacharparenright}{\isacharparenright}}\\
\isa{eval{\isacharunderscore}expr\ \freeify{t}\ {\isacharparenleft}\constructor{Binop}\ \freeify{f}\ \freeify{e\isactrlisub {\isadigit{1}}}\ \freeify{e\isactrlisub {\isadigit{2}}}{\isacharparenright}} & \isa{\holkeyword{let}\ {\isacharparenleft}\boundify{D\isactrlisub {\isadigit{1}}}{\isacharcomma}\ \boundify{f\isactrlisub {\isadigit{1}}}{\isacharparenright}\ {\isacharequal}\ eval{\isacharunderscore}expr\ \freeify{t}\ \freeify{e\isactrlisub {\isadigit{1}}}{\isacharsemicolon}\isanewline
\isaindent{\holkeyword{let}\ }{\isacharparenleft}\boundify{D\isactrlisub {\isadigit{2}}}{\isacharcomma}\ \boundify{f\isactrlisub {\isadigit{2}}}{\isacharparenright}\ {\isacharequal}\ eval{\isacharunderscore}expr\ {\isacharparenleft}\freeify{t}\ {\isacharplus}\ used{\isacharunderscore}tmps\ \freeify{e\isactrlisub {\isadigit{1}}}{\isacharparenright}\ \freeify{e\isactrlisub {\isadigit{2}}}\isanewline
\holkeyword{in}\ {\isacharparenleft}\boundify{D\isactrlisub {\isadigit{1}}}\ {\isasymunion}\ \boundify{D\isactrlisub {\isadigit{2}}}{\isacharcomma}\ {\isasymlambda}\boundify{{\isasymtheta}}{\isachardot}\ \freeify{f}\ {\isacharparenleft}\boundify{f\isactrlisub {\isadigit{1}}}\ \boundify{{\isasymtheta}}{\isacharparenright}\ {\isacharparenleft}\boundify{f\isactrlisub {\isadigit{2}}}\ \boundify{{\isasymtheta}}{\isacharparenright}{\isacharparenright}}
\end{tabularx}
\caption{Expression evaluation\label{fig:expression-evaluation}}
\end{figure}
The function \isa{used{\isacharunderscore}tmps\ \freeify{e}} calculates the number of temporaries that are necessary to evaluate expression \isa{\freeify{e}}, where every \isa{\constructor{Mem}} expression accounts to one temporary. 
With \isa{issue{\isacharunderscore}expr\ \freeify{t}\ \freeify{e}} we obtain the instruction list for expression \isa{\freeify{e}} starting at temporary \isa{\freeify{t}}, whereas \isa{eval{\isacharunderscore}expr\ \freeify{t}\ \freeify{e}} constructs the operation as a pair of the domain and a function on the temporaries.

The program transitions are defined in Figure~\ref{fig:program-transitions}. We instantiate the program state by a tuple \isa{{\isacharparenleft}\freeify{s}{\isacharcomma}\ \freeify{t}{\isacharparenright}} containing the statement \isa{\freeify{s}} and the temporary counter \isa{\freeify{t}}.
\begin{figure}
\begin{center}
\isa{\mbox{}\inferrule{\mbox{{\isasymforall}\boundify{sop}{\isachardot}\ \freeify{a}\ {\isasymnoteq}\ \constructor{Tmp}\ \boundify{sop}}\\\ \mbox{\freeify{a{\isacharprime}}\ {\isacharequal}\ \constructor{Tmp}\ {\isacharparenleft}eval{\isacharunderscore}expr\ \freeify{t}\ \freeify{a}{\isacharparenright}}\\\ \mbox{\freeify{t{\isacharprime}}\ {\isacharequal}\ \freeify{t}\ {\isacharplus}\ used{\isacharunderscore}tmps\ \freeify{a}}\\\ \mbox{\freeify{is}\ {\isacharequal}\ issue{\isacharunderscore}expr\ \freeify{t}\ \freeify{a}}}{\mbox{\freeify{{\isasymtheta}}{\isasymturnstile}\ {\isacharparenleft}\constructor{Assign}\ \freeify{volatile}\ \freeify{a}\ \freeify{e}\ \freeify{A}\ \freeify{L}\ \freeify{R}\ \freeify{W}{\isacharcomma}\ \freeify{t}{\isacharparenright}\ {\isasymrightarrow}\isactrlsub p\ {\isacharparenleft}{\isacharparenleft}\constructor{Assign}\ \freeify{volatile}\ \freeify{a{\isacharprime}}\ \freeify{e}\ \freeify{A}\ \freeify{L}\ \freeify{R}\ \freeify{W}{\isacharcomma}\ \freeify{t{\isacharprime}}{\isacharparenright}{\isacharcomma}\ \freeify{is}{\isacharparenright}}}}\\[0.5\baselineskip]
\isa{\mbox{}\inferrule{\mbox{\freeify{D}\ {\isasymsubseteq}\ dom\ \freeify{{\isasymtheta}}}\\\ \mbox{\freeify{is}\ {\isacharequal}\ issue{\isacharunderscore}expr\ \freeify{t}\ \freeify{e}\ {\isacharat}\ {\isacharbrackleft}\constructor{Write}\ \freeify{volatile}\ {\isacharparenleft}\freeify{a}\ \freeify{{\isasymtheta}}{\isacharparenright}\ {\isacharparenleft}eval{\isacharunderscore}expr\ \freeify{t}\ \freeify{e}{\isacharparenright}\ {\isacharparenleft}\freeify{A}\ \freeify{{\isasymtheta}}{\isacharparenright}\ {\isacharparenleft}\freeify{L}\ \freeify{{\isasymtheta}}{\isacharparenright}\ {\isacharparenleft}\freeify{R}\ \freeify{{\isasymtheta}}{\isacharparenright}\ {\isacharparenleft}\freeify{W}\ \freeify{{\isasymtheta}}{\isacharparenright}{\isacharbrackright}}}{\mbox{\freeify{{\isasymtheta}}{\isasymturnstile}\ {\isacharparenleft}\constructor{Assign}\ \freeify{volatile}\ {\isacharparenleft}\constructor{Tmp}\ {\isacharparenleft}\freeify{D}{\isacharcomma}\ \freeify{a}{\isacharparenright}{\isacharparenright}\ \freeify{e}\ \freeify{A}\ \freeify{L}\ \freeify{R}\ \freeify{W}{\isacharcomma}\ \freeify{t}{\isacharparenright}\ {\isasymrightarrow}\isactrlsub p\ {\isacharparenleft}{\isacharparenleft}\constructor{Skip}{\isacharcomma}\ \freeify{t}\ {\isacharplus}\ used{\isacharunderscore}tmps\ \freeify{e}{\isacharparenright}{\isacharcomma}\ \freeify{is}{\isacharparenright}}}}\\[0.5\baselineskip]
\isa{\mbox{}\inferrule{\mbox{{\isasymforall}\boundify{sop}{\isachardot}\ \freeify{a}\ {\isasymnoteq}\ \constructor{Tmp}\ \boundify{sop}}\\\ \mbox{\freeify{a{\isacharprime}}\ {\isacharequal}\ \constructor{Tmp}\ {\isacharparenleft}eval{\isacharunderscore}expr\ \freeify{t}\ \freeify{a}{\isacharparenright}}\\\ \mbox{\freeify{t{\isacharprime}}\ {\isacharequal}\ \freeify{t}\ {\isacharplus}\ used{\isacharunderscore}tmps\ \freeify{a}}\\\ \mbox{\freeify{is}\ {\isacharequal}\ issue{\isacharunderscore}expr\ \freeify{t}\ \freeify{a}}}{\mbox{\freeify{{\isasymtheta}}{\isasymturnstile}\ {\isacharparenleft}\constructor{CAS}\ \freeify{a}\ \freeify{c\isactrlisub e}\ \freeify{s\isactrlisub e}\ \freeify{A}\ \freeify{L}\ \freeify{R}\ \freeify{W}{\isacharcomma}\ \freeify{t}{\isacharparenright}\ {\isasymrightarrow}\isactrlsub p\ {\isacharparenleft}{\isacharparenleft}\constructor{CAS}\ \freeify{a{\isacharprime}}\ \freeify{c\isactrlisub e}\ \freeify{s\isactrlisub e}\ \freeify{A}\ \freeify{L}\ \freeify{R}\ \freeify{W}{\isacharcomma}\ \freeify{t{\isacharprime}}{\isacharparenright}{\isacharcomma}\ \freeify{is}{\isacharparenright}}}}\\[0.5\baselineskip]
\isa{\mbox{}\inferrule{\mbox{{\isasymforall}\boundify{sop}{\isachardot}\ \freeify{c\isactrlisub e}\ {\isasymnoteq}\ \constructor{Tmp}\ \boundify{sop}}\\\ \mbox{\freeify{c\isactrlisub e{\isacharprime}}\ {\isacharequal}\ \constructor{Tmp}\ {\isacharparenleft}eval{\isacharunderscore}expr\ \freeify{t}\ \freeify{c\isactrlisub e}{\isacharparenright}}\\\ \mbox{\freeify{t{\isacharprime}}\ {\isacharequal}\ \freeify{t}\ {\isacharplus}\ used{\isacharunderscore}tmps\ \freeify{c\isactrlisub e}}\\\ \mbox{\freeify{is}\ {\isacharequal}\ issue{\isacharunderscore}expr\ \freeify{t}\ \freeify{c\isactrlisub e}}}{\mbox{\freeify{{\isasymtheta}}{\isasymturnstile}\ {\isacharparenleft}\constructor{CAS}\ {\isacharparenleft}\constructor{Tmp}\ \freeify{a}{\isacharparenright}\ \freeify{c\isactrlisub e}\ \freeify{s\isactrlisub e}\ \freeify{A}\ \freeify{L}\ \freeify{R}\ \freeify{W}{\isacharcomma}\ \freeify{t}{\isacharparenright}\ {\isasymrightarrow}\isactrlsub p\ {\isacharparenleft}{\isacharparenleft}\constructor{CAS}\ {\isacharparenleft}\constructor{Tmp}\ \freeify{a}{\isacharparenright}\ \freeify{c\isactrlisub e{\isacharprime}}\ \freeify{s\isactrlisub e}\ \freeify{A}\ \freeify{L}\ \freeify{R}\ \freeify{W}{\isacharcomma}\ \freeify{t{\isacharprime}}{\isacharparenright}{\isacharcomma}\ \freeify{is}{\isacharparenright}}}}\\[0.5\baselineskip]
\isa{\mbox{}\inferrule{\mbox{\freeify{D\isactrlisub a}\ {\isasymsubseteq}\ dom\ \freeify{{\isasymtheta}}}\\\ \mbox{\freeify{D\isactrlisub c}\ {\isasymsubseteq}\ dom\ \freeify{{\isasymtheta}}}\\\ \mbox{eval{\isacharunderscore}expr\ \freeify{t}\ \freeify{s\isactrlisub e}\ {\isacharequal}\ {\isacharparenleft}\freeify{D}{\isacharcomma}\ \freeify{f}{\isacharparenright}}\\\ \mbox{\freeify{t{\isacharprime}}\ {\isacharequal}\ \freeify{t}\ {\isacharplus}\ used{\isacharunderscore}tmps\ \freeify{s\isactrlisub e}}\\\ \mbox{\freeify{cond}\ {\isacharequal}\ {\isacharparenleft}{\isasymlambda}\boundify{{\isasymtheta}}{\isachardot}\ the\ {\isacharparenleft}\boundify{{\isasymtheta}}\ \freeify{t{\isacharprime}}{\isacharparenright}\ {\isacharequal}\ \freeify{c}\ \boundify{{\isasymtheta}}{\isacharparenright}}\\\ \mbox{\freeify{ret}\ {\isacharequal}\ {\isacharparenleft}{\isasymlambda}\boundify{v\isactrlisub {\isadigit{1}}}\ \boundify{v\isactrlisub {\isadigit{2}}}{\isachardot}\ \boundify{v\isactrlisub {\isadigit{1}}}{\isacharparenright}}\\\ \mbox{\freeify{is}\ {\isacharequal}\ issue{\isacharunderscore}expr\ \freeify{t}\ \freeify{s\isactrlisub e}\ {\isacharat}\ {\isacharbrackleft}\constructor{RMW}\ {\isacharparenleft}\freeify{a}\ \freeify{{\isasymtheta}}{\isacharparenright}\ \freeify{t{\isacharprime}}\ {\isacharparenleft}\freeify{D}{\isacharcomma}\ \freeify{f}{\isacharparenright}\ \freeify{cond}\ \freeify{ret}\ {\isacharparenleft}\freeify{A}\ \freeify{{\isasymtheta}}{\isacharparenright}\ {\isacharparenleft}\freeify{L}\ \freeify{{\isasymtheta}}{\isacharparenright}\ {\isacharparenleft}\freeify{R}\ \freeify{{\isasymtheta}}{\isacharparenright}\ {\isacharparenleft}\freeify{W}\ \freeify{{\isasymtheta}}{\isacharparenright}{\isacharbrackright}}}{\mbox{\freeify{{\isasymtheta}}{\isasymturnstile}\ {\isacharparenleft}\constructor{CAS}\ {\isacharparenleft}\constructor{Tmp}\ {\isacharparenleft}\freeify{D\isactrlisub a}{\isacharcomma}\ \freeify{a}{\isacharparenright}{\isacharparenright}\ {\isacharparenleft}\constructor{Tmp}\ {\isacharparenleft}\freeify{D\isactrlisub c}{\isacharcomma}\ \freeify{c}{\isacharparenright}{\isacharparenright}\ \freeify{s\isactrlisub e}\ \freeify{A}\ \freeify{L}\ \freeify{R}\ \freeify{W}{\isacharcomma}\ \freeify{t}{\isacharparenright}\ {\isasymrightarrow}\isactrlsub p\ {\isacharparenleft}{\isacharparenleft}\constructor{Skip}{\isacharcomma}\ Suc\ \freeify{t{\isacharprime}}{\isacharparenright}{\isacharcomma}\ \freeify{is}{\isacharparenright}}}}\\[0.5\baselineskip]
\isa{\mbox{}\inferrule{\mbox{\freeify{{\isasymtheta}}{\isasymturnstile}\ {\isacharparenleft}\freeify{s\isactrlisub {\isadigit{1}}}{\isacharcomma}\ \freeify{t}{\isacharparenright}\ {\isasymrightarrow}\isactrlsub p\ {\isacharparenleft}{\isacharparenleft}\freeify{s\isactrlisub {\isadigit{1}}{\isacharprime}}{\isacharcomma}\ \freeify{t{\isacharprime}}{\isacharparenright}{\isacharcomma}\ \freeify{is}{\isacharparenright}}}{\mbox{\freeify{{\isasymtheta}}{\isasymturnstile}\ {\isacharparenleft}\constructor{Seq}\ \freeify{s\isactrlisub {\isadigit{1}}}\ \freeify{s\isactrlisub {\isadigit{2}}}{\isacharcomma}\ \freeify{t}{\isacharparenright}\ {\isasymrightarrow}\isactrlsub p\ {\isacharparenleft}{\isacharparenleft}\constructor{Seq}\ \freeify{s\isactrlisub {\isadigit{1}}{\isacharprime}}\ \freeify{s\isactrlisub {\isadigit{2}}}{\isacharcomma}\ \freeify{t{\isacharprime}}{\isacharparenright}{\isacharcomma}\ \freeify{is}{\isacharparenright}}}}\\[-0.3\baselineskip] 
\isa{\mbox{}\inferrule{\mbox{}}{\mbox{\freeify{{\isasymtheta}}{\isasymturnstile}\ {\isacharparenleft}\constructor{Seq}\ \constructor{Skip}\ \freeify{s\isactrlisub {\isadigit{2}}}{\isacharcomma}\ \freeify{t}{\isacharparenright}\ {\isasymrightarrow}\isactrlsub p\ {\isacharparenleft}{\isacharparenleft}\freeify{s\isactrlisub {\isadigit{2}}}{\isacharcomma}\ \freeify{t}{\isacharparenright}{\isacharcomma}\ {\isacharbrackleft}{\isacharbrackright}{\isacharparenright}}}} \\[0.5\baselineskip]
\isa{\mbox{}\inferrule{\mbox{{\isasymforall}\boundify{sop}{\isachardot}\ \freeify{e}\ {\isasymnoteq}\ \constructor{Tmp}\ \boundify{sop}}\\\ \mbox{\freeify{e{\isacharprime}}\ {\isacharequal}\ \constructor{Tmp}\ {\isacharparenleft}eval{\isacharunderscore}expr\ \freeify{t}\ \freeify{e}{\isacharparenright}}\\\ \mbox{\freeify{t{\isacharprime}}\ {\isacharequal}\ \freeify{t}\ {\isacharplus}\ used{\isacharunderscore}tmps\ \freeify{e}}\\\ \mbox{\freeify{is}\ {\isacharequal}\ issue{\isacharunderscore}expr\ \freeify{t}\ \freeify{e}}}{\mbox{\freeify{{\isasymtheta}}{\isasymturnstile}\ {\isacharparenleft}\constructor{Cond}\ \freeify{e}\ \freeify{s\isactrlisub {\isadigit{1}}}\ \freeify{s\isactrlisub {\isadigit{2}}}{\isacharcomma}\ \freeify{t}{\isacharparenright}\ {\isasymrightarrow}\isactrlsub p\ {\isacharparenleft}{\isacharparenleft}\constructor{Cond}\ \freeify{e{\isacharprime}}\ \freeify{s\isactrlisub {\isadigit{1}}}\ \freeify{s\isactrlisub {\isadigit{2}}}{\isacharcomma}\ \freeify{t{\isacharprime}}{\isacharparenright}{\isacharcomma}\ \freeify{is}{\isacharparenright}}}}\\[0.5\baselineskip]
\isa{\mbox{}\inferrule{\mbox{\freeify{D}\ {\isasymsubseteq}\ dom\ \freeify{{\isasymtheta}}}\\\ \mbox{isTrue\ {\isacharparenleft}\freeify{e}\ \freeify{{\isasymtheta}}{\isacharparenright}}}{\mbox{\freeify{{\isasymtheta}}{\isasymturnstile}\ {\isacharparenleft}\constructor{Cond}\ {\isacharparenleft}\constructor{Tmp}\ {\isacharparenleft}\freeify{D}{\isacharcomma}\ \freeify{e}{\isacharparenright}{\isacharparenright}\ \freeify{s\isactrlisub {\isadigit{1}}}\ \freeify{s\isactrlisub {\isadigit{2}}}{\isacharcomma}\ \freeify{t}{\isacharparenright}\ {\isasymrightarrow}\isactrlsub p\ {\isacharparenleft}{\isacharparenleft}\freeify{s\isactrlisub {\isadigit{1}}}{\isacharcomma}\ \freeify{t}{\isacharparenright}{\isacharcomma}\ {\isacharbrackleft}{\isacharbrackright}{\isacharparenright}}}}\\[0.5\baselineskip]
\isa{\mbox{}\inferrule{\mbox{\freeify{D}\ {\isasymsubseteq}\ dom\ \freeify{{\isasymtheta}}}\\\ \mbox{{\isasymnot}\ isTrue\ {\isacharparenleft}\freeify{e}\ \freeify{{\isasymtheta}}{\isacharparenright}}}{\mbox{\freeify{{\isasymtheta}}{\isasymturnstile}\ {\isacharparenleft}\constructor{Cond}\ {\isacharparenleft}\constructor{Tmp}\ {\isacharparenleft}\freeify{D}{\isacharcomma}\ \freeify{e}{\isacharparenright}{\isacharparenright}\ \freeify{s\isactrlisub {\isadigit{1}}}\ \freeify{s\isactrlisub {\isadigit{2}}}{\isacharcomma}\ \freeify{t}{\isacharparenright}\ {\isasymrightarrow}\isactrlsub p\ {\isacharparenleft}{\isacharparenleft}\freeify{s\isactrlisub {\isadigit{2}}}{\isacharcomma}\ \freeify{t}{\isacharparenright}{\isacharcomma}\ {\isacharbrackleft}{\isacharbrackright}{\isacharparenright}}}}\\[-0.3\baselineskip]
\isa{\mbox{}\inferrule{\mbox{}}{\mbox{\freeify{{\isasymtheta}}{\isasymturnstile}\ {\isacharparenleft}\constructor{While}\ \freeify{e}\ \freeify{s}{\isacharcomma}\ \freeify{t}{\isacharparenright}\ {\isasymrightarrow}\isactrlsub p\ {\isacharparenleft}{\isacharparenleft}\constructor{Cond}\ \freeify{e}\ {\isacharparenleft}\constructor{Seq}\ \freeify{s}\ {\isacharparenleft}\constructor{While}\ \freeify{e}\ \freeify{s}{\isacharparenright}{\isacharparenright}\ \constructor{Skip}{\isacharcomma}\ \freeify{t}{\isacharparenright}{\isacharcomma}\ {\isacharbrackleft}{\isacharbrackright}{\isacharparenright}}}}\\[-0.3\baselineskip]
\isa{\mbox{}\inferrule{\mbox{}}{\mbox{\freeify{{\isasymtheta}}{\isasymturnstile}\ {\isacharparenleft}\constructor{SGhost}\ \freeify{A}\ \freeify{L}{\isacharcomma}\ \freeify{t}{\isacharparenright}\ {\isasymrightarrow}\isactrlsub p\ {\isacharparenleft}{\isacharparenleft}\constructor{Skip}{\isacharcomma}\ \freeify{t}{\isacharparenright}{\isacharcomma}\ {\isacharbrackleft}\constructor{Ghost}\ {\isacharparenleft}\freeify{A}\ \freeify{{\isasymtheta}}{\isacharparenright}\ {\isacharparenleft}\freeify{L}\ \freeify{{\isasymtheta}}{\isacharparenright}{\isacharbrackright}{\isacharparenright}}}}\\[-0.3\baselineskip]
\isa{\mbox{}\inferrule{\mbox{}}{\mbox{\freeify{{\isasymtheta}}{\isasymturnstile}\ {\isacharparenleft}\constructor{SFence}{\isacharcomma}\ \freeify{t}{\isacharparenright}\ {\isasymrightarrow}\isactrlsub p\ {\isacharparenleft}{\isacharparenleft}\constructor{Skip}{\isacharcomma}\ \freeify{t}{\isacharparenright}{\isacharcomma}\ {\isacharbrackleft}\constructor{Fence}{\isacharbrackright}{\isacharparenright}}}}\\[0.1\baselineskip]
\end{center}
\caption{Program transitions\label{fig:program-transitions}}
\end{figure}
To assign an expression \isa{\freeify{e}} to an address(-expression) \isa{\freeify{a}} we first create the memory instructions for evaluation the address \isa{\freeify{a}} and transforming the expression to an operation on temporaries. The temporary counter is incremented accordingly. 
When the value is available in the temporaries we continue by
creating the memory instructions for evaluation of expression \isa{\freeify{e}} followed by the corresponding store operation.
Note that the ownership annotations can depend on the temporaries and thus can take the calculated address into account.

Execution of compare and swap \isa{\constructor{CAS}} involves evaluation of three expressions, the address \isa{\freeify{a}} the compare value \isa{\freeify{c\isactrlisub e}} and the swap value \isa{\freeify{s\isactrlisub e}}. 
It is finally mapped to the read-modify-write instruction \isa{\constructor{RMW}} of the memory system. 
Recall that execution of \isa{\constructor{RMW}} first stores the memory content at address \isa{\freeify{a}} to the specified temporary. 
The condition compares this value with the result of evaluating \isa{\freeify{c\isactrlisub e}} and writes swap value \isa{\freeify{s\isactrlisub a}} if successful. 
In either case the temporary finally returns the old value read.
 
Sequential composition is straightforward. An if-then-else is computed by first issuing the memory instructions for evaluation of condition \isa{\freeify{e}} and transforming the condition to an operation on temporaries. 
When the result is available the transition to the first or second statement is made, depending on the result of \isa{isTrue}.
Execution of the loop is defined by stepwise unfolding.
Ghost and fence statements are just propagated to the memory system.
To instantiate Theorem~\ref{thm:simulation} with PIMP we define the invariant parameter \isa{\freeify{valid}}, which has to be maintained by all transitions of PIMP, the memory system and the store buffer. 
Let \isa{\freeify{{\isasymtheta}}} be the valuation of temporaries in the current configuration, for every thread configuration \isa{\freeify{ts\isactrlisub s\isactrlisub b}\ensuremath{_{[\freeify{i}]}}\ {\isacharequal}\ {\isacharparenleft}{\isacharparenleft}\freeify{s}{\isacharcomma}\ \freeify{t}{\isacharparenright}{\isacharcomma}\ \freeify{is}{\isacharcomma}\ \freeify{{\isasymtheta}}{\isacharcomma}\ \freeify{sb}{\isacharcomma}\ \freeify{{\isasymD}}{\isacharcomma}\ \freeify{{\isasymO}}{\isacharcomma}\ \freeify{{\isasymA}}{\isacharparenright}} where \isa{\freeify{i}\ {\isacharless}\ {\isacharbar}\freeify{ts\isactrlisub s\isactrlisub b}{\isacharbar}} we require:
\begin{inparaenum}
\item The domain of all intermediate \isa{\constructor{Tmp}\ {\isacharparenleft}\freeify{D}{\isacharcomma}\ \freeify{f}{\isacharparenright}} expressions in statement \isa{\freeify{s}} is below counter \isa{\freeify{t}}.
\item All temporaries in the memory system including the store buffer are below counter \isa{\freeify{t}}.
\item All temporaries less than counter \isa{\freeify{t}} are either already defined in the temporaries \isa{\freeify{{\isasymtheta}}} or are outstanding read temporaries in the memory system.
\end{inparaenum}

For the PIMP transitions we prove these invariants by rule induction on the semantics.
For the memory system (including the store buffer steps) the invariants are straightforward. 
The memory system does not alter the program state and does not create new temporaries, only the PIMP transitions create new ones in strictly  ascending order.%
\end{isamarkuptext}%
\isamarkuptrue%
\isadelimtheory
\endisadelimtheory
\isatagtheory
\endisatagtheory
{\isafoldtheory}%
\isadelimtheory
\endisadelimtheory
\end{isabellebody}%

\section{Conclusion \label{sec:conclusion}}

We have presented a practical and flexible concurrent programming
discipline that ensures sequential consistency on TSO machines, such
as present in current x64 architectures. Our approach covers a wide
variety of concurrency control, covering locking, data races, single
writer multiple readers, read only and thread local portions of
memory.  We minimize the need for store buffer flushes to optimize the
usage of the hardware.  Our theorem is not coupled to a specific
logical framework like separation logic but is based on more
fundamental arguments, namely the adherence to the access and flushing
policy which can be discharged within any program logic.

\paragraph{Related work.}

A categorization of various weak memory models is presented in
\cite{Adve:Computer-29-12-66}.  It is
compatible with the recent revisions of the Intel manuals
\cite{Intel:IIA2006-ALL} and the revised x86 model presented in
\cite{Owens:TPHOL09-?}.  The state of the art in formal verification
of concurrent programs is still based on a sequentially consistent
memory model.  To justify this on a weak memory model often a quite
drastic approach is chosen, allowing only coarse-grained concurrency
usually implemented by locking. Thereby data races are ruled out
completely and there are results that data race free programs can be
considered as sequentially consistent for example for the Java memory
model \cite{DBLP:conf/ecoop/SevcikA08,DBLP:conf/tphol/AspinallS07} or
the x86 memory model\cite{Owens:TPHOL09-?}.  Ridge
\cite{conf/tphol/Ridge07} considers weak memory and data-races and
verifies Peterson's mutual exclusion algorithm. He ensures
sequentially consistency by flushing after every write to shared
memory.
Burckhardt and Musuvathi\cite{Sober} describe an execution monitor that 
efficiently checks whether a sequentially consistent TSO execution has a single-step
extension that is not sequentially consistent. Like our approach, it
avoids having to consider the store buffers as an explicit part of the
state. However, their condition requires maintaining in ghost state
enough history information to determine causality between events,
which means maintaining a vector clock (which is itself unbounded) for
each memory address. Moreover, causality (being essentially graph
reachability) is already not first-order, and hence unsuitable for
many types of program verification. 

\paragraph{Future work.}
We currently have an asymmetry in the ghost operations for ownership
transfer: whereas we have a `free flowing' ghost operation to acquire
an address which can appear anywhere, releases are delayed to the next
(volatile or interlocked) write operation.  As sketched in Section
\ref{sec:simulation} delaying releases is motivated by our simulation
proof.  To rule out certain races we argue that the state of the
virtual machine is at most more restrictive as the state of the store
buffer machine.  As the virtual state is obtained by flushing all
store buffers until the first volatile write is hit (excluding it) a
release in that flushed section would violate this invariant. However,
we believe this does not restrict expressibility as the only way for
other threads to gain knowledge about the released address is via the
next volatile write operation of the thread.  Formally we want to
liberate the points where an ownership release can happen by
introducing free flowing releases.  The informal key argument is that
an unsafe execution with delayed releases implies an unsafe execution
with free flowing releases.  Between the release point in the free
flowing execution and the delayed release point (at the next volatile
write) a thread only executes commuting operations with respect to bad
races.  Such a race at the delayed point already justifies a race at
the release point in the free flowing execution.

Another direction of future work is to take compiler optimization into
account. Our volatile accesses correspond roughly to volatile memory
accesses within a C program. An optimizing compiler is free to convert
any sequence of non-volatile accesses into a (sequentially
semantically equivalent) sequence of accesses. As long as execution is
sequentially consistent, equivalence of these programs (\eg with
respect to final states of executions that end with volatile
operations) follows immediately by reduction. However, some compilers
are a little more lenient in their optimizations, and allow operations
on certain local variables to move across volatile operations. In the
context of C (where pointers to stack variables can be passed by
pointer), the notion of ``locality'' is somewhat tricky, and makes
essential use of C forbidding (semantically) address arithmetic across
memory objects.

Finally, we should note that there are important programs that, in the
presence of store buffers, are correct but not sequentially
consistent. A typical example is the following simplified form of
barrier synchronization: each processor has a flag that it writes and
other processors read, and each processor waits for all processors to
set their flags before continuing past the barrier. This is not
sequentially consistent -- each processor might see his own flag set
and later see all other flags clear -- but it is still correct. One
possibility is to give a more general reduction theorem that allows
each processor to always treat store buffers of other processors as
empty, and its own store buffer as empty except for brief periods of
time.

\section*{Acknowledgements}
We thank Mark Hillebrand for discussions and feedback on this work and
extensive comments on this report.

\bibliographystyle{plain}
\bibliography{storebuffer}

\end{document}